\documentclass[useAMS,usenatbib]{mn2e}
\usepackage{times,psfig,epsfig}

\def\lsim{\,\lower2truept\hbox{${<\atop\hbox{\raise4truept\hbox{$\sim$}}}$}\,}
\def\gsim{\,\lower2truept\hbox{${> \atop\hbox{\raise4truept\hbox{$\sim$}}}$}\,}

\author{A. Bonaldi, S. Ricciardi, S. Leach, F. Stivoli, C. Baccigalupi, G. De Zotti}
\title[WMAP 3yr data with the CCA]{WMAP 3yr data with the CCA: anomalous emission
and impact of component separation on the CMB power spectrum}
\begin{document}
\maketitle
\begin{abstract}
The Correlated Component Analysis (CCA) allows us to estimate
how the different diffuse emissions mix in CMB experiments, taking also
into account complementary information from other surveys. It is
especially useful to deal with possible additional components for
which little or no prior information exists. An application of CCA
to WMAP maps assuming that only the canonical Galactic emissions
(synchrotron, free-free, and thermal dust) are present, highlights
the widespread presence of a spectrally flat ``synchrotron''
component, largely uncorrelated with the synchrotron template,
suggesting that an additional foreground is indeed required. We have
tested various spectral shapes for such component, namely a power
law as expected if it is flat synchrotron, and two spectral shapes
that may fit the spinning dust emission: a parabola in the $\log
S$--$\log \nu$ plane, and a grey body. If the spatial distribution
of the additional (``anomalous'') component is not constrained a
priori it is found to be always tightly correlated with thermal
dust, but the correlation is not perfect.  Quality tests applied to
the reconstructed CMB maps clearly disfavour two of the models. 
The CMB power spectra, estimated from CMB maps reconstructed 
exploiting the
three surviving foreground models, are generally consistent with the
WMAP ones, although at least one of them gives a significantly
higher quadrupole moment than found by the WMAP team. Taking foreground 
modeling uncertainties into account, we find that the mean quadrupole 
amplitude for the three ``good'' models is
less than $1\sigma$ below the expectation from the standard
$\Lambda$CDM model. Also the other reported deviations from model
predictions are found not to be statistically significant, except
for the excess power at $l\simeq
40$. 
We confirm the evidence for a marked north south asymmetry in
the large scale ($l<20$) CMB anisotropies, which is stable with
respect to the foreground parametrization we adopted.
We also present a first, albeit preliminary, all sky map 
of the ``anomalous emission''.

\end{abstract}
\section{Introduction}

The \emph{Wilkinson Microwave Anisotropy Probe} \citep[WMAP,][]{bennett2003,spergel2006} data are
having a dominant role in determining the reference cosmological
model. It is therefore important to get an in depth understanding of
all the uncertainties in the data analysis and of their effects on
the estimates of the angular power spectrum of the Cosmic Microwave
Background (CMB), from which the values of the fundamental
cosmological parameters are derived.

In this paper we present a new investigation, exploiting a novel
technique, of the effect of foreground contamination. CMB power
spectrum analyses are normally done removing a substantial fraction
of the sky ($\simeq 25\%$) around the Galactic plane where Galactic
emissions are stronger. However, the uncertainties in the underlying 
foreground emission make it possible that a relevant contribution from 
them remains, even when large sky cuts are considered. \cite{slosarseljak} 
developed a formalism to
incorporate foreground uncertainties and the effect of sky cuts in
the estimates of errors on multipole moments. \cite{slosar2004} marginalized over the foreground templates produced
by \cite{bennett2003} from WMAP data. \cite{odwyer2005} and \cite{odwyer2004} have shown that it is possible to include the
foreground components in a self-consistent fashion within the
statistical framework of their approach to power spectrum
estimation, based on a Bayesian statistical analysis of the data
utilizing Gibbs Sampling.

Here we deal with foregrounds applying to the three-year WMAP data
the Correlated Component Analysis
\citep[CCA,][]{bedini2005,bonaldi2006}, which exploits a second
order statistics of the data to estimate the spectral behaviour of
the mixed components directly from the WMAP data. The method also
allows us to incorporate the additional information from other
surveys, such as the \cite{haslam1982} 408 MHz map, the IRAS/COBE
dust emission map \citep{finkbeiner1999}, and the extinction
corrected H$\alpha$ map \citep{dickinson2003}, as a tracer of
free-free emission. At variance with most alternative techniques,
CCA allows us to perform an essentially all-sky component
separation: only a narrow strip around the Galactic equator ($|b| <
3^\circ$, $\simeq 5\%$ of the sky) needs to be removed \citep{bonaldi2006}.

Of course, we need to specify a priori which foreground components
are contributing to the maps and describe their frequency spectra
with a small number of parameters, as specified is \S\,2, and this
brings in the issue of the possible presence of one more Galactic
component, in addition to the well established synchrotron,
free-free, and thermal dust emissions. In \S\,3 we describe the
estimation with CCA of the mixing matrix, exploited to reconstruct
the maps of astrophysical components using an inversion algorithm
that, in our case, is the harmonic Wiener Filter, as described in
\S\,4. The properties of the reconstructed foreground components are
discussed in \S\,5, and those of the CMB in \S\,6 where we analyze
the uncertainties on the derived CMB power spectrum ensuing from
those on foreground properties, especially in relation to the
reported deviations from predictions of the ``concordance''
cosmological model \citep{spergel2006}. The main conclusions are
summarized in \S\,7.



\section{Diffuse components in the WMAP bands}\label{sec:foreg}

We summarize here the frequency spectra of diffuse components, in
terms of antenna temperature. The CMB blackbody spectrum is:
\begin{equation}
T_{\rm A,CMB}(\nu)=\frac{(h\nu/kT_{\rm CMB})^2\exp (h\nu/kT_{\rm
CMB})}{(\exp (h\nu/kT_{\rm CMB})-1)^2} \label{cmb}
\end{equation}
where $h$ and $k$ are the Planck and the Boltzmann constant,
respectively, and $T_{\rm CMB}=2.726\,$K.

The free-free emission spectrum is also quite uniform and well
approximated, in the frequency range of interest, by:
\begin{equation}
T_{\rm A,ff}(\nu)\propto \nu^{-2.14} \label{ff}
\end{equation}
It is difficult to obtain a spatial template of this emission since
synchrotron dominates radio surveys and thermal dust dominates
far-IR surveys. A good tracer of free-free emission are $H\alpha$
maps, but they need to be corrected for dust extinction, and this
correction adds a significant uncertainty \citep{dickinson2003}.

The synchrotron spectrum is usually modeled as:
\begin{equation}
T_{\rm A,synch}(\nu)\propto \nu^{-\beta_s}. \label{syn}
\end{equation}
Its spectral index, $\beta_s$, is known to vary across the sky,
reflecting the variations of the energy index of relativistic
electrons \citep{rybicki1979}. The WMAP team
\citep{bennett2003,hinshaw2006} find steeper values ($\beta_s \simeq
3$) at high Galactic latitudes and flatter values ($\beta_s \simeq
2.5$) close to star-forming regions, especially in the Galactic
plane.  The mean value is 2.65 between 408 MHz and the K-band, and
2.69 between 408 MHz and the Q-band. A spectral steepening is
expected at high microwave frequencies, as a consequence of electron
ageing effects \citep{banday1991}. The \cite{haslam1982} 408 MHz
map is commonly used as a synchrotron template.

The spectrum of the thermal (vibrational) emission from dust grains
in the interstellar medium is conveniently described by a grey body:
\begin{equation}
T_{\rm A,dust} (\nu)\propto \frac{\nu^{\beta_d+1}}{\exp
(h\nu/kT_{\rm dust})-1} \label{dust}
\end{equation}
with $T_{\rm dust}\sim 18\,$K and $\beta_d \sim 1.67$ at mm
wavelengths \citep{finkbeiner1999}. Both $T_{\rm
dust}$ and $\beta_d$ are expected to vary across the sky due to
variations of the interstellar radiation field and of the dust
grains composition. Measurements of $\beta_d$ generally lie in the
range $1.5 \leq \beta_d \leq 2.0$, but an even broader range has
been observed by the PRONAOS experiment \citep{dupac2003}: $0.8 \leq
\beta_d \leq 2.4$. The WMAP data only weakly constrain the dust
spectral index \citep{hinshaw2006}. The most frequently used
spatial template of thermal dust emission is the one worked out by
\cite{finkbeiner1999} combining IRAS and COBE maps.

In recent years evidence has been reported for an additional,
thermal dust correlated component called ``anomalous microwave
emission'', that may dominate, at least in some Galactic regions, in
the 20--40 GHz range, where it has a spectrum similar to
synchrotron. It may be originated by tiny, fast spinning, dust
grains \citep{DL1998,oliveira1999,oliveira2002}. Indications of this
emission were found by analyses of CMB experiments like Saskatoon
\citep{oliveira1997}, 19 GHz \citep{oliveira1998}, Tenerife
\citep{oliveira1999,oliveira2002,muk2001}; Python V \citep{muk2003}.
Although \cite{bennett2003} concluded that spinning dust emission
contributes less than 5\% of the WMAP Ka-band antenna temperature,
evidence of the anomalous emission were uncovered combining WMAP
data with other measurements, especially at lower frequencies
\citep{lagache2003,oliveira2004,finkbeiner2004,watson2005,davies2006}.
Moreover, the polarization properties of the dust-correlated
low-frequency component differ from those of the component well
correlated with the Haslam template, suggesting a different emission
mechanism \citep{page2006}. Nevertheless, the exact nature, the
spectral properties, and the spatial distribution of this foreground
remain uncertain. There is not general agreement even on its
existence; a combination of free-free emission and of strongly self
absorbed synchrotron could also account for the data
\citep{hinshaw2006}.

Point sources are another important foreground component, but their
contribution to the power spectrum is important only on scales
$\lsim 1^\circ$, while in this analysis we are mostly interested on
larger scales, where diffuse emissions dominate. For the present
analysis we will simply need to mask the brightest sources, as
described in the following sections.



\begin{table}
\caption{WMAP channels} \label{tab:wmap}
\begin{tabular}{|c|c|c|c|c|c|}
\hline
Channel&K&Ka&Q&V&W\\
\hline
Frequency (GHz)&23&33&41&61&94\\
\hline
Resolution (degrees)&0.93&0.68&0.53&0.35&$<$0.23\\
\hline
$\sigma_0$(mK)&1.42&1.45&2.21&3.11&6.50\\
\hline
mean pixel rms (mK)&0.07&0.07&0.07&0.09&0.1\\
\hline
\end{tabular}
\end{table}
\section{Mixing matrix estimation with CCA}\label{sec:ccawork}

\subsection{Outline of the method}\label{sec:cca}
As usual, we express the data vector $\mathbf{x}$ in each pixel as:
\begin{equation}
\mathbf{x}=\mathbf{H}\mathbf{s}+\mathbf{n}\label{model}
\end{equation}
where  $\mathbf{H}$ is a $M \times N$ mixing matrix, $\mathbf{s}$ is
the $N$-vector of sources (components) and  $\mathbf{n}$ the
$M$-vector of instrumental noise ($M$ is then the number of
independent maps used in the analysis). The generic element $h_{\rm
dc}$ of the mixing matrix is proportional to the spectrum of the
c-th source at an effective frequency within the d-th sensor
passband. Using eq.~(\ref{model}) we implicitly assume that the
spatial pattern of the physical processes is independent of
frequency and that the effects of the telescope beam has been
equalized in all channels.

Given a generic signal $\mathbf{X}$, defined in a two dimensional
space with coordinates $(\xi,\eta)$, the covariance matrix of this
signal is:
\begin{eqnarray}
\mathbf{C}_X(\tau,\psi)=\langle
[\mathbf{X(\tau,\psi)}-\mu][\mathbf{X}(\xi+\tau,\eta+\psi)-\mu]^T
\rangle
\end{eqnarray}
where $\langle ... \rangle$ denotes expectation under the
appropriate joint probability, $\mu$ is the mean vector and the
superscript $T$ means transposition. Every covariance matrix is
characterized by the shift pair $(\tau,\psi)$, where $\tau$ and
$\psi$ are increments in the $\xi$ and $\eta$ coordinates.

From eq.~(\ref{model}) we can easily derive a relation between the
data covariance matrix $\mathbf{C}_x$ at a certain lag, the source
covariance matrix $\mathbf{C}_s$ at the same lag, the mixing matrix
$\mathbf{H}$, and the noise covariance matrix $\mathbf{C}_n$:
\begin{equation}
\mathbf{C}_x(\tau,\psi)=\mathbf{H}\mathbf{C}_s(\tau,\psi)\mathbf{H}^T+\mathbf{C}_n(\tau,\psi).\label{problem}
\end{equation}
The covariance matrix $\mathbf{C}_x$ can be estimated from the data
as:
\begin{equation}
\hat{\mathbf{C}_x} (\tau,\psi) =\frac{1}{N_p} \sum_{\xi,
\eta}[\mathbf{x}(\xi,\eta)-\mu _x
][\mathbf{x}(\xi+\tau,\eta+\phi)-\mu _x ]^T \label{covariance}
\end{equation}
where $N_p$ is the number of pixels sampling the data. Given a noise
process, we can model the noise correlation matrix $\mathbf{C}_n$:
for example, if noise can be assumed signal-independent, white and
zero-mean, for $(\tau,\psi)=(0,0)$ $\mathbf{C}_n$ is a diagonal
matrix whose elements are the noise variances in the measured
channels, while for  $(\tau,\psi)\neq (0,0)$ $\mathbf{C}_n$ is the
null $M \times M$ matrix. If the noise process deviates
significantly from this ideal model, $\mathbf{C}_n$ can be computed
using Monte Carlo on noise maps in the same way we did for $\mathbf{C}_x$.

Once $\mathbf{C}_x$ and $\mathbf{C}_n$ are known,
eq.~(\ref{problem}) can be used to identify the mixing operator
$\mathbf{H}$. The strategy of CCA is to parameterize the mixing
matrix to reduce the number of unknowns and to take into account
enough nonzero shift pairs $(\tau,\psi)$ to estimate both
$\mathbf{H}$ and $\mathbf{C}_s$. To solve the identification problem
we perform the minimization:
\begin{eqnarray}
(\Gamma(:,:))=\hbox{argmin}\sum _{\tau,\psi}\parallel \mathbf{H}(\Gamma) \mathbf{C}_s [\Sigma(\tau,\psi)] \mathbf{H}^T(\Gamma)+\\
-\hat{\mathbf{C}_x}(\tau,\psi)- \mathbf{C}_n (\tau,\psi) \parallel
\nonumber
\end{eqnarray}
where $\Gamma$ is the vector of all parameters defining $\mathbf{H}$
and $\Sigma(:,:)$ is the vector containing all the unknown elements
of matrices $\mathbf{C}_s$ for every shift pair.

The main product of CCA is an estimate of the mixing matrix
$\mathbf{H}$. The result is defined apart from the normalization 
of the different signals at a given frequency. Therefore  we
estimate a normalized mixing matrix, obtained by assigning value 1
to all the elements of an arbitrarily chosen row. This matrix can be
used to perform the source reconstruction with standard inversion
methods. In our case, this is done by harmonic Wiener Filtering, as
we will describe in \S~\ref{sec:wf}.

\subsection{Input data}\label{sec:input}
The basic data are the three-year WMAP maps in the K, Ka, Q, V and W
bands, whose main characteristics are summarized in Table
\ref{tab:wmap}. The values of $\sigma_0$ reported for every channel
allow us to calculate a noise map at each frequency as
$\sigma=\sigma_0 N_{\rm obs}^{-1/2}$, where $N_{\rm obs}$ is the map
of the effective number of observations, provided with the data.
The mean rms per pixel relative to each channel at full resolution 
is also reported.  

The maps were then smoothed to the common resolution of $1^\circ$
\citep{bennett2003}. Since we do not have an exact formula to
associate an rms level to each map after smoothing, we therefore simulated 
a set of ten WMAP noise maps at each WMAP
frequency using the corresponding $\sigma_0$ and $N_{\rm obs}$. We then
smoothed them to the $1^\circ$ resolution, and computed the mean rms for each channel.

We assumed the noise process to be gaussian with variance
equal to the mean of the all-sky variances of the simulated noise
maps. Then $\mathbf{C}_n(\tau,\psi)$ is a diagonal matrix whose
elements are these empirical variances for $(\tau,\psi)=(0,0)$, and
the null matrix for $(\tau,\psi)\neq(0,0)$. In eq.~(\ref{problem})
we then treat $\mathbf{C}_n(\tau,\psi)$ as a diagonal matrix whose
elements are these empirical variances for $(\tau,\psi)=(0,0)$, and
as null matrices for $(\tau,\psi)\neq(0,0)$.

Strong point sources can contribute substantially to the mean
surface brightness in some pixels and, since their frequency spectra
are generally different from those of diffuse components, they may
bias the mixing matrix estimates. Source fluxes are highly diluted
in the relatively low resolution maps used here and we find that the
point source effects become negligible once we mask a region of
$1^\circ$ radius around each source contained in the three-year
point source catalogue provided with the WMAP data. In addition,
based on the analysis by \cite{bonaldi2006}, we applied a Galactic
cut of $\pm 3^\circ$. Finally, we masked few highly contaminated
regions, namely CenA, the Large Magellanic Cloud, $\rho$ Oph, Orion A, Orion B and Tau A.

To get a better leverage for foreground differentiation we
complemented the WMAP data with a ``thermal dust'' and a
``synchrotron'' map. The former was obtained extrapolating to 850
GHz the dust map by \cite{schlegel1998} using the best fit model of
\cite{finkbeiner1999}. The latter is based on the \cite{haslam1982}
a 408 MHz map. Even if the Haslam map is often assumed to be pure
synchrotron, contributions from the free-free emission can be important, 
especially on the Galactic plane \citep{paladini2005}.
Our ``synchrotron'' map was then obtained by subtracting from the
Haslam map the free-free contribution estimated from the $H\alpha$
map corrected for dust absorption \citep{dickinson2003}.

\subsection{Analysis}\label{sec:mat}
To reduce the number of unknowns to a manageable level we need to
parameterize the mixing matrix $\mathbf{H}$. This means we have to
exploit our knowledge of foregrounds to model the WMAP data in terms
of a small number of parameters. On the other hand, as mentioned in
\S\,1, we are not even sure about the number of foreground
components that need to be taken into account.



Also, foreground spectral parameters vary across the sky. Therefore
we would ideally want a map of the values of such parameters with
sufficiently high resolution. However, CCA needs a large number
of independent pixels to derive them. As discussed by \cite{bonaldi2006} a good compromise is to apply CCA to patches of about
$1500\,\hbox{deg}^2$. We worked with patches of $(\Delta l,\Delta
b)=(50^\circ, 30^\circ)$ on the Galactic plane, and increased
$\Delta l$ with increasing $|b|$ as necessary to roughly preserve
the pixel number (and the patch area). We centered our patches at
longitudes $l_c =\{0^\circ,40^\circ, 80^\circ,... 320^\circ \}$ and
latitudes $b_c = \{0^\circ,\pm 15^\circ,\pm 25^\circ,\pm
35^\circ\}$.

The different sky patches intersect each other,
but the overlap is small enough for them to be taken as independent
of each other. In fact, the covariance matrices
[eq.~(\ref{covariance})] of adjacent patches are all substantially
different from each other.

For any given input model CCA has then yielded the mixing matrix
for each of the 63 patches. As discussed in the following
subsection, the first use of these results was to give us some
guidance in the selection of input models.
In this analysis and throughout the whole paper we used the HEALPix
tessellation scheme \citep{gorski2005} \footnote{see http://healpix.jpl.nasa.gov}.


\subsubsection{The standard foreground model, {\bf M1}}\label{p:M1}
We started our analysis adopting the most conservative model (which
will be referred to as {\bf M1}), containing the standard mixture of
CMB, synchrotron, thermal dust and free-free. This model has only
one free parameter, the synchrotron spectral index $\beta_s$. In
fact, the spectral behaviour of the CMB and of the free-free
emissions are known [eqs.~(\ref{cmb}) and ~(\ref{ff})], and the
thermal dust spectrum is only very poorly constrained by WMAP data,
so that the results are very weakly dependent on the assumed dust
temperature and emissivity index. We adopted the commonly used
values $T_{\rm dust}=18\,$K and $\beta_d=1.67$ [eq.~(\ref{dust})].

\begin{figure}
\begin{center}
\includegraphics[width=5cm,angle=90]{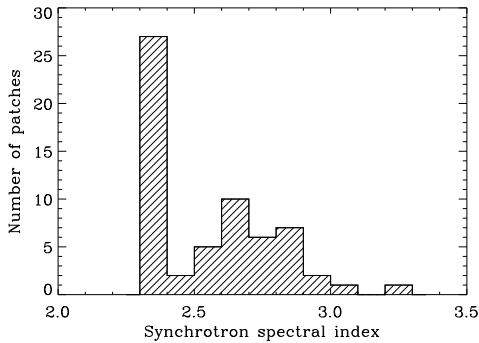}
\caption{Distribution of the synchrotron spectral index $\beta_s$
obtained by CCA for model {\bf M1}.}
\label{fig:isto_M1}
\end{center}
\end{figure}

The distribution of $\beta_s$ obtained with CCA, shown in
Fig. \ref{fig:isto_M1}, presents two peaks. To determine position and width of the peaks we divided the distribution in two, $\beta_s \le 2.4$ and $\beta_s > 2.4$, and fitted them with gaussian distributions. One peak is at $\beta_s
\simeq 2.7$ and has a dispersion $\sigma \simeq 0.2$, which is roughly
what is expected for synchrotron. The second one, at $\beta_s
=2.3852$, is extremely narrow $\sigma \simeq 0.0004$, hinting at a
different component. We explicitly checked that this distribution is
unaffected by different choices of the thermal dust parameters,
within the observed ranges (\S\,2). This immediately begs the question, what is this second flatter component. 

\begin{figure}
\begin{center}
\includegraphics[width=5cm,angle=90]{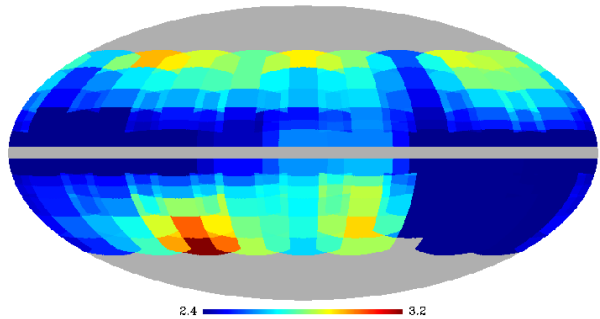}
\caption{Map of the synchrotron spectral indices recovered by the
CCA for model {\bf M1}.  }
\label{fig:validzones}
\end{center}
\end{figure}

In Fig.~\ref{fig:validzones} we show the map of the recovered
synchrotron spectral indices. Mean values are adopted in the regions
where different patches overlap. The flattish component is mainly 
located at low latitudes, more than $\sim
40^\circ$ away from the Galactic center, and is not well correlated
with the synchrotron template. In the next sub-section we
investigated different possible spectral shapes for this
``anomalous'' component.

\subsection{Models for the ``anomalous'' component}
\begin{figure}
\begin{center}
\includegraphics[width=5cm,angle=90]{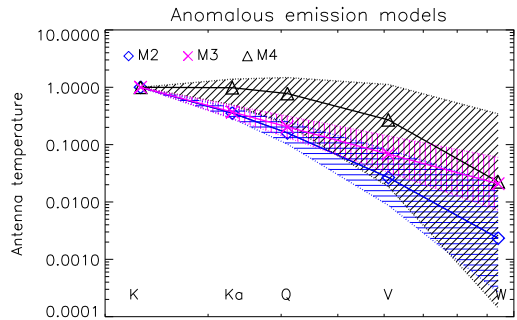}
\caption{Frequency spectra of the anomalous emission for the models
{\bf M2}, {\bf M3} and {\bf M5}. The curves are normalized at the K
band. For each model, the shaded area shows the range of values
allowed for the esitmation, while the symbols refer to the central values.}
\label{fig:anomalous_models}
\end{center}
\end{figure}

As pointed out by many authors (e.g. de Oliveira-Costa et al. 2004;
Watson et al. 2005), the spectral behaviour of the anomalous
emission clearly differs from those of free-free and synchrotron at
10--15 GHz (where we don't have all-sky maps), but is similar to
them in the 20--40 GHz range. This means that any attempt to
estimate simultaneously the spectral parameters of synchrotron and
of the anomalous emission from WMAP data is liable to strong
aliasing effects. We have therefore chosen to estimate the spectral
parameters of the anomalous emission keeping $\beta_s$ fixed, and
repeating the analysis for several values of it in the range $2.8
\leq \beta_s \leq 3.1$. Specifically we ran CCA over all sky
patches for $\beta_{s} =\{2.80, 2.86, 2.92, 2.98, 2.04, 3.10\}$.

As for the anomalous emission spectrum, we have investigated the
models summarized in Table~\ref{tab:models} with their free
parameters, and described in the following. The proposed frequency scaling for the anomalous emission are also shown in Fig.~\ref{fig:anomalous_models}.

\begin{table*}
\caption{Summary of the models} \label{tab:models}

\begin{tabular}{|c|c|c|c|}
\hline
Model&investigated component&parametrization&free parameters and ranges\\
\hline
{\bf{M1}}&Synchrotron& eq.~(\protect\ref{syn})& $2.0 \leq \beta_s \leq 3.5$\\
{\bf{M2}}&Anomalous emission& eq.~(\protect\ref{tspin}); $\nu_{\rm max}=20\,$GHz &  $1.0 \leq m_{60} \leq 5.0$\\
{\bf{M3}}&Anomalous emission&$T_{\rm A,X}(\nu)\propto \nu^{-\beta_2}$ &$2.0 \leq \beta_2 \leq 2.6$\\
{\bf{M4}}&Anomalous emission&$T_{\rm A,X} (\nu)\propto [\nu^{\beta_x+1}]/[\exp (h\nu/kT_{x})-1]$;\, $\beta_x=2.4$ \,
& $0.2\,\hbox{K} \leq T_x \leq 0.7\,\hbox{K}$ \\
{\bf{M5}}&Thermal + Anomalous dust& eq.~(\protect\ref{spindust});
$\nu_{\rm max}=20\,$GHz
& $1.0 \leq m_{60} \leq 5.0$, \, $0.1 \leq R_{60} \leq 2.0$ \\
\hline
\end{tabular}
\end{table*}

\subsubsection{Model {\bf M2}}
The results by \cite{oliveira2004}, \cite{watson2005}, and
\cite{davies2006} suggest that the spectrum of the anomalous
emission for $\nu > 20\,$GHz may be represented by a parabola in the
($\log \nu , \log S$) plane with $\nu_{\rm max}=20\,$GHz:

\begin{eqnarray}
\log T_{\rm A,X}(\nu) = \hbox{const} + \left(\frac{-m_{60}\log
\nu_{\rm max}}{\log(\nu_{\rm max}/60\,\hbox{GHz})}-2\right)\log \nu + \nonumber \\ 
 -\frac{-m_{60}}{2\log(\nu_{\rm max}/60\,\hbox{GHz})}(\log
\nu)^2 \label{tspin}.
\end{eqnarray}
with $\nu$ in GHz. The free parameter, $m_{60}$, is the angular coefficient  at the frequency of 60 GHz in the ($\log \nu , \log S$) plane.

\begin{figure}
\begin{center}
\includegraphics[width=5.5cm,angle=90]{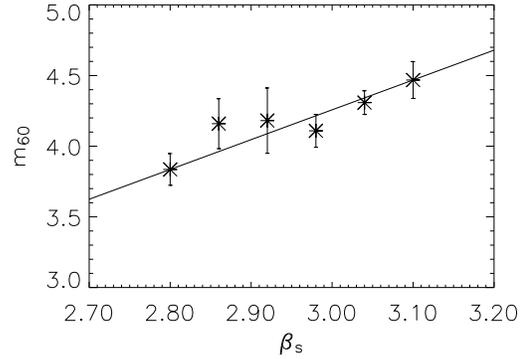}

\caption{Mean estimated values of $m_{60}$, with their errors,  vs
assumed values of the synchrotron spectral index $\beta_s$ (model
{\bf M2}).}

\label{fig:isto_M2}
\end{center}
\end{figure}

Running CCA for the above set of values for
$\beta_{s}$ we found that the mean values of $m_{60}$ over the sky
patches are in the range $3.8 \leq m_{60} \leq 4.5$, and correlate
with $\beta_{s}$. The linear best fit relation, shown in
Fig.~\ref{fig:isto_M2}, is:
\begin{equation}
m_{60}=(2.1101\pm0.0005) \beta_s -(2.073\pm 0.002) \label{m2}.
\end{equation}
The corresponding spectral shape is compatible with the anomalous emission 
detected by \cite{davies2006}. In particular, our scaling between the K and 
Ka bands is almost the same they find, and the one  between Ka and Q is very 
similar.
Our model falls down more rapidly then theirs at higher frequencies, 
but is still inside their error bars.

\subsubsection{Model {\bf M3}}
Alternatively, the anomalous emission might be interpreted as
flat-spectrum synchrotron, possibly highly self absorbed, associated
to strong magnetic fields local to star forming regions (and thus
dust-correlated). To investigate this possibility we parameterize
this component as a power law, with a spectral index $\beta_2$.

As before, we obtained the distribution of $\beta_{2}$ over the sky
patches for each value of $\beta_{s}$. Such distribution turned out
to be quite narrow (dispersion $\sigma =  0.003$) and independent of
$\beta_s$. The mean value of the spectral index
\begin{equation}
\langle\beta_2\rangle =2.144 \label{m3}
\end{equation}
turns out to be remarkably close to that of free-free, hinting at
the possibility of aliasing effects. We will come back to this
possibility in \S\,\ref{sec:sep}.

\subsubsection{Model {\bf M4}}
In this case we adopt for the anomalous component the spectrum used
by \cite{tegmark2000}: it is a grey body, of the form of eq.~(\ref{dust}),
with temperature $T_x = 0.25\,$K and emissivity index $\beta_x =
2.4$. We checked that a simultaneous estimation of both $\beta_x$
and $T_x$ was not feasible, as the algorithm failed to converge.
Then, we fixed $\beta_x=2.4$ and allowed $T_x$ to vary in the range
$0.2\,\hbox{K} \leq T_x \leq 0.7\,\hbox{K}$. 

As it can be seen looking at Fig.~\ref{fig:anomalous_models}, 
{\bf M2} and {\bf M4} yield quite different frequency scalings. 
Infact, the latter 
has been proposed to fit typical \cite{DL1998} spinning
dust models, but it is not suited to reproduce the results 
by \cite{davies2006}.
Given the slope
 between the K and Ka bands, {\bf M4} goes down more rapidly.

\begin{figure}
\begin{center}
\includegraphics[width=5.5cm,angle=90]{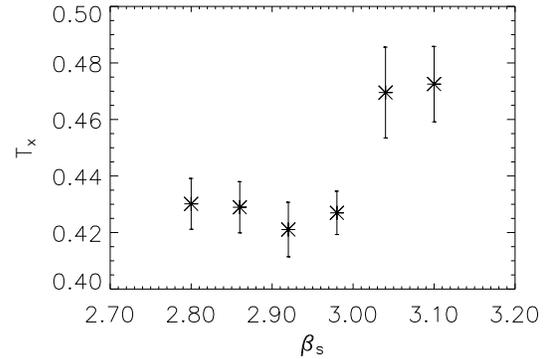}

\caption{Mean estimated values of $T_x$, with their errors,  vs
assumed values of the synchrotron spectral index $\beta_s$ (model
{\bf M4}).}

\label{fig:isto_M4}
\end{center}
\end{figure}

The results (Fig.~\ref{fig:isto_M4}) can be summarized as:
\begin{equation}
T_x=  \Big\{\begin{array}{ll}
0.43 \pm 0.01\,\hbox{K} & \hbox{if}\ \beta_s \le 3.0\\
0.47 \pm 0.01\,\hbox{K} & \hbox{if}\  \beta_s > 3.0. \end{array}
\label{m4}
\end{equation}
The spectrum rises from K to Ka band, and slowly decreases going to higher 
frequencies. The slope between 60 and 94 GHz is consistent with what found 
for {\bf M2}.

\subsubsection{Model {\bf M5}}
We tested the possibility that there is a perfect correlation
between thermal dust and anomalous emission, so that they can be
treated as a single component parameterized as:
\begin{equation}
T_{\rm A,dust+X}(\nu)\propto T_{\rm A,dust}(\nu) +R_{60}\cdot T_{\rm
A,X}(\nu)\label{spindust},
\end{equation}
where $T_{\rm A,dust}$ is given by eq.~(\ref{dust}) and $T_{\rm
A,X}$ by eq.~(\ref{tspin}). The factor $R_{60}$ quantifies the relative
intensity of the two emissions at a frequency of 60 GHz.  
The free parameters of this model
are $R_{60}$ and $m_{60}$, whereas the other ones are fixed as in {\bf
M2}.


\begin{figure}
\begin{center}
\includegraphics[width=5.5cm,angle=90]{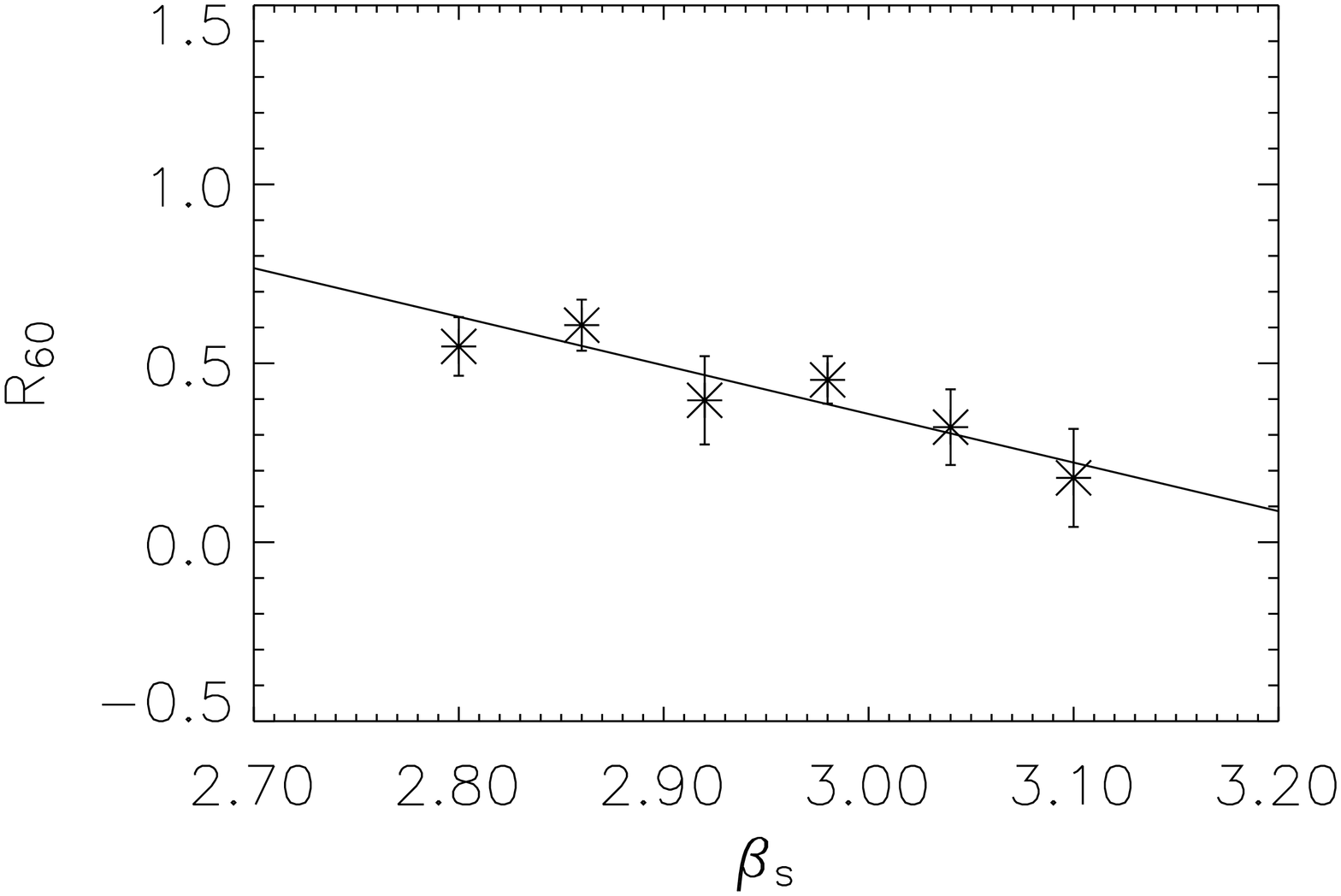}
\includegraphics[width=5.5cm,angle=90]{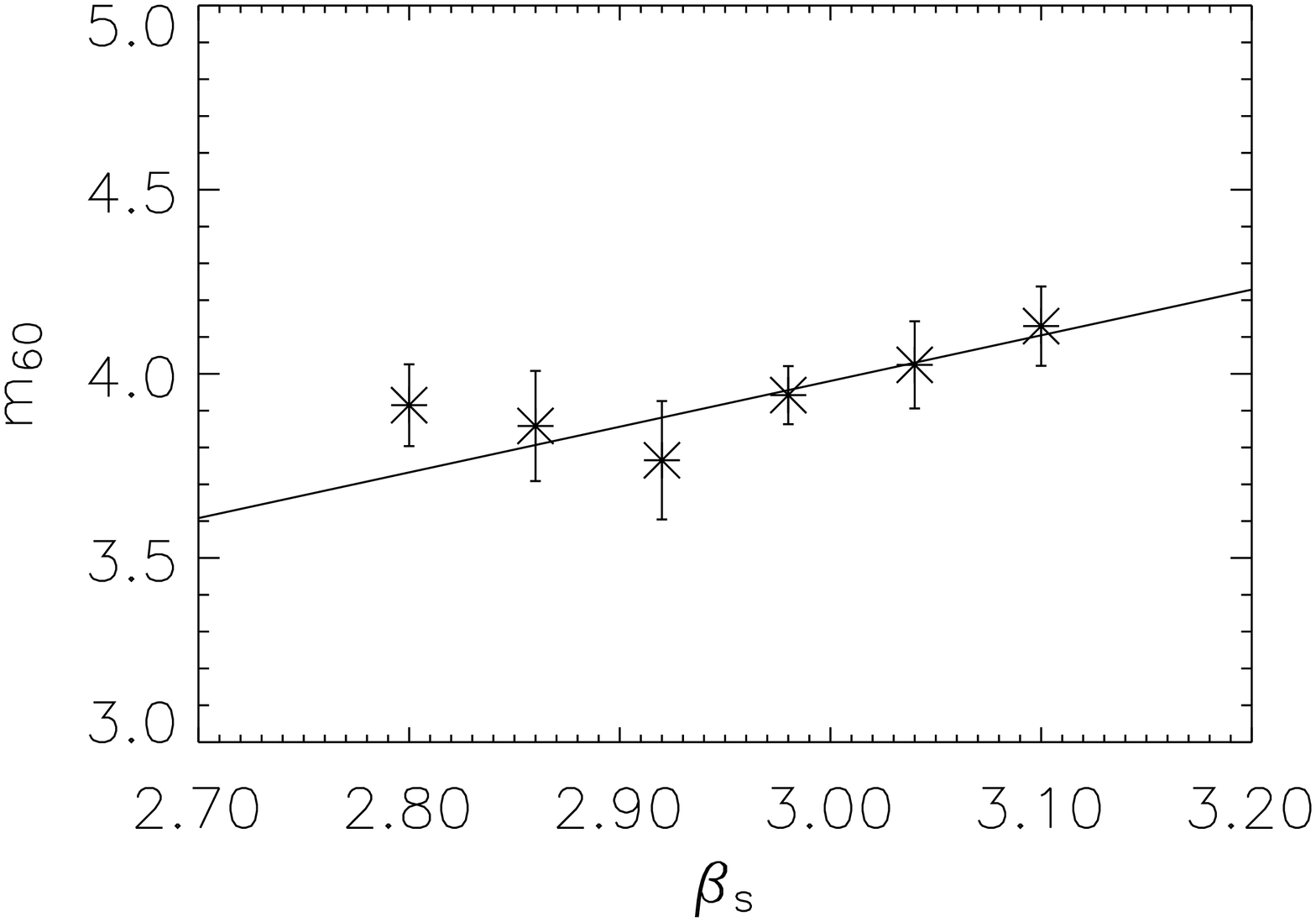}

\caption{Mean estimated values of $R_{60}$ and $m_{60}$, with their 
errors, vs assumed values of the synchrotron spectral index
$\beta_s$ (model {\bf M5}).} \label{fig:isto_M5}
\end{center}
\end{figure}

The recovered values of the parameters are shown in Fig.~\ref{fig:isto_M5}. The values of $m_{60}$ are within the range found
for model {\bf M2}, which did not assume any correlation with
thermal dust, but the correlation with $\beta_s$ is now weaker. The
linear best fit relations of $m_{60}$ and $R_{60}$ with $\beta_s$ (solid
lines in Fig. \ref{fig:isto_M5}) are:
\begin{eqnarray}
R_{60}=(-1.4 \pm 0.4)\beta_s+(4.4 \pm 1.2)\label{m5_1}\\
m_{60}=(1.3 \pm 0.3)\beta_s +(0.3 \pm 0.1).\label{m5_2}
\end{eqnarray}
This is again consistent with \cite{davies2006}, both for 
the slope $m_{60}$ and the intensity ratio of the anomalous component respect 
to thermal dust $R_{60}$.

\section{Source reconstruction with the WF}\label{sec:wf}
Once the mixing matrices of the diffuse components in the WMAP data
were estimated for the different models, we performed the source
reconstruction with the harmonic Wiener Filtering. In the harmonic
space the problem stated in eq.~(\ref{model}) simply becomes:
\begin{equation}
\mathbf{x}_{lm}=\mathbf{H}\mathbf{s}_{lm}+\mathbf{n}_{lm}
\label{model2},
\end{equation}
where the vectors $\mathbf{x}_{lm}$, $\mathbf{s}_{lm}$, and
$\mathbf{n}_{lm}$ contain the harmonic coefficients of channels,
sources and instrumental noise, respectively. 
Using a linear approach to component separation, an estimate of the vector $\mathbf{s}_{lm}$, say $\mathbf{\hat{s}%
}_{lm}$, can be obtained as:
\begin{equation}
\mathbf{\hat
s}_{lm}=\mathbf{W}_{\hat{H}}^{(l)}\mathbf{x}_{lm}\label{recon},
\end{equation}
where the reconstruction matrix $\mathbf{W}_{\hat{H}}^{(l)}$ depends
on the estimate $\mathbf{\hat{H}}$ of the mixing matrix. In the case
of Wiener Filtering, we have:
\begin{equation}
\mathbf{W}_{\hat{H}}^{(l)}
=\mathbf{C_{s}}^{(l)}\mathbf{\hat{H}}^{T}[\mathbf{{\hat{H}}C_{s}}^{(l)}
\mathbf{\hat{H}}^{T} +\mathbf{C_{n}}^{(l)}]^{-1},  \label{26}
\end{equation}
where $\mathbf{C_{s}}$ and $\mathbf{C_{n}}$ are the source and the
noise power spectra, respectively. If the source and the noise
processes are uncorrelated, $\mathbf{C_{s}}$ and
$\mathbf{C_{n}}$  are diagonal matrices. 
Even in this approximation, however,  reconstructing the components
with WF requires the knowledge of the diagonal elements
$C_{s}^{(l)}(i) \equiv \mathbf{C_{s}}^{(l)}(i,i)$ and
$C_{n}^{(l)}(i) \equiv \mathbf{C_{n}}^{(l)}(i,i)$. The latter can be
easily modeled; in the case of uniform noise, we have:
\begin{equation}
{C_{n}}^{(l)}(i)=4\pi \sigma _{i}^{2}/N_{i},  \label{30}
\end{equation}%
where $N_{i}$ is the number of pixels, and  $\sigma_{i}$ is the rms
pixel noise. On the other hand, the power spectra of the components,
$C_{s}^{(l)}$, are not, or only poorly, known a priori.  A common
approach \citep{hobson1998, stolyarov2002} is to start from a rough
estimate of the source power spectra $C_{s}^{(l)}(i)$ and get the
reconstruction matrix through eq.~(\ref{26}). The latter allows us
to get the  estimated components through eq.~(\ref{recon}). After
the first reconstruction, we obtain $C_{s}^{(l)}(i)$ as the unbiased
estimators of the power spectra of the reconstructed sources and we
update $\mathbf{W}_{\hat{H}}^{(l)}$. The procedure is iterated until
convergence on the power spectra is reached.


To derive the initial power spectra, we performed a first,
low-resolution, reconstruction through eq.~(\ref{recon}) exploiting
the reconstruction matrix:
\begin{equation}
\mathbf{W}_{\hat{H}}^{(l)}
=(\mathbf{\hat{H}}^{T}(\mathbf{C_{n}}^{(l)})^{-1}\mathbf{\hat{H}})^{-1}
\mathbf{\hat{H}}^{T}(\mathbf{C_{n}}^{(l)})^{-1},
\end{equation}
which corresponds to a maximum likelihood analysis. Note that in
this case, besides the estimate of the mixing matrix
$\mathbf{\hat{H}}$, recovered by CCA, only the prior knowledge
of the noise power spectra $\mathbf{C_{n}}^{(l)}$ is required. We
then use as initial $C_{s}^{(l)}(i)$ for the WF the power spectra of
these reconstructed sources extrapolated to small scales assuming $l^{-3}$ behaviour \citep{Zavarise}. After convergence has been reached, dust, free-free and, if present, anomalous emission behave as $l^{-2}$ between multipoles $\sim 10$ and few hundreds, while synchrotron behaves as $l^{-3}$.

The approach just described uses a minimum number of priors: in
practice, the components are identified only by their frequency
scalings. The lack of prior covariance information certainly limits
the quality of the reconstruction, but in this way we avoid biasing
the results.

\begin{figure}
\begin{center}
\includegraphics[width=5.cm,angle=90]{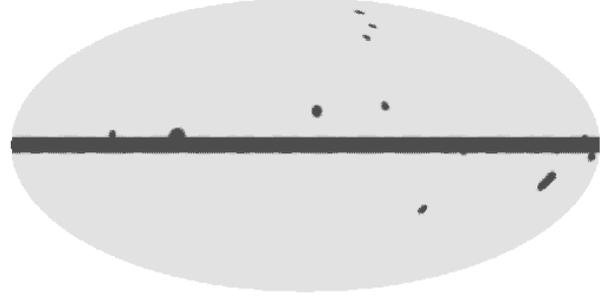}

\caption{Apodized mask used to preprocess the input maps for WF. The
minimum (dark grey) and maximum (light grey) values are 0 and 1
respectively.}

\label{fig:apo0}
\end{center}
\end{figure}

\subsection{The WF data set}\label{sec:wfinput}
For the matrix inversion we used
the WMAP  K, Ka, Q, V, and W bands at their original resolution.
Since we now work in the harmonic domain, a frequency-dependent
beamwidth can be naturally accounted for.  We assumed circularly
symmetric gaussian beams with the FWHM values reported in
Table~\ref{tab:wmap}. Even if this may not be a good approximation
of the real beams, the effect of beam asymmetries is unlikely to be
relevant on the relatively large angular scales we are considering
here. The noise processes have been assumed to be uniform over the
whole sky: to compute the noise levels we averaged the standard
deviations of a set of simulated WMAP noise maps.

To take advantage of all the available information we complemented
the WMAP data with the synchrotron, dust and free-free maps adopted
for the mixing matrix estimation (see \S\,\ref{sec:ccawork}). These
maps have resolutions of $60'$, $5'$ and $60'$, respectively. We
have allowed for the effect of variations of spectral indices of
synchrotron and dust across the sky (while the inversion is done
adopting everywhere the mean values obtained with CCA) by
attributing to the corresponding maps uncertainties much larger than
the measurements errors. In particular, we computed the standard
deviation of each prior map outside the apodized mask of Fig.
\ref{fig:apo0} and assumed a constant rms level
of 20\% for the  408 MHz synchrotron
map and of 10\% for the 850 GHz thermal dust map. The spread of
spectral indices may induce even larger uncertainties, but we have
checked that increasing the uncertainties by up to a factor of two has
a negligible impact on the reconstruction of the CMB and a small
effect on that of foregrounds. This problem does not affect the
free-free for which we have adopted the nominal rms error of 7\%
\citep{dickinson2003}.

Algorithms working in the spherical harmonics space require data
maps defined over the whole sky. On the other hand, cutting out
highly contaminated regions is generally necessary to recover the
CMB map. But the spherical harmonic basis is no longer orthonormal
over the cut sphere and correlations are introduced between harmonic
coefficients at different ($l$, $m$) modes. As shown by
\cite{stolyarov2005}, one way to overcome these problems is to
consider the cut areas as an extreme case of anisotropic noise,
where the noise becomes formally infinite. In this way we can work
with all sky maps but the solution is not constrained in the omitted
areas. Similar results are obtained, with less computational
problems, by filling the masked areas with a noise realization at
the nominal level for the considered channel (Stolyarov, private
communication). To reduce edge effects the maps were weighted with
the apodized mask shown in Fig.~\ref{fig:apo0}. This mask excludes
the strip at $|b|<3^\circ$, the LMC, the Cen A region, and regions
of $1^\circ$ radius centered on the brightest point sources.
Obviously, all pixels which have been modified will be excluded from
the analysis of the output maps.

\subsection{Analysis}\label{sec:wfanalysis}
The CCA gives, for each model, the distribution of spectral
parameters found in different sky patches. These distribution
reflect (besides estimation errors) the spatial variability of the
spectra. On the other hand, the harmonic Wiener Filtering is applied
using a single mixing matrix over the full sky, i.e. assuming
spatially uniform spectra. To investigate the effect of this
simplification we generated a set of 30 mixing matrices for each
model, drawing the spectral parameters at random from CCA
distributions.
For each matrix, we performed a full-sky source
reconstruction up to $l=300$.

In the cases of models {\bf M2} and {\bf M5}, we
adopted, for $\beta_s$, the same distribution of {\bf M1}, and derived the value of $m_{60}$ or of $R_{60}$ and
$m_{60}$ from eq. ~(\ref{m2}) or from eqs.~(\ref{m5_1}) and
(\ref{m5_2}), respectively. In the case of {\bf M3}, we used the
same uniform distribution for $\beta_s$ and a Gaussian distribution
for $\beta_2$. For {\bf M4} we used the
same uniform distribution for $\beta_s$ and derived $T_x$ according to eq~(\ref{m4}).

A particular case is given by the standard foreground model, {\bf
M1}. The results obtained with CCA for this model indicate that,
if no additional component is present, we cannot assume a uniform
synchrotron spectral index all over the sky. Thus, generating a set
of synchrotron spectral indices according to the distribution shown
in Fig.~\ref{fig:isto_M1} would not be meaningful. For this reason
we used this model for reference, adopting for the synchrotron
spectral index a uniform distribution in the range $2.8 \le \beta_s
\le 3.1$. 

\begin{figure*}
\begin{center}
\includegraphics[width=12cm,angle=90]{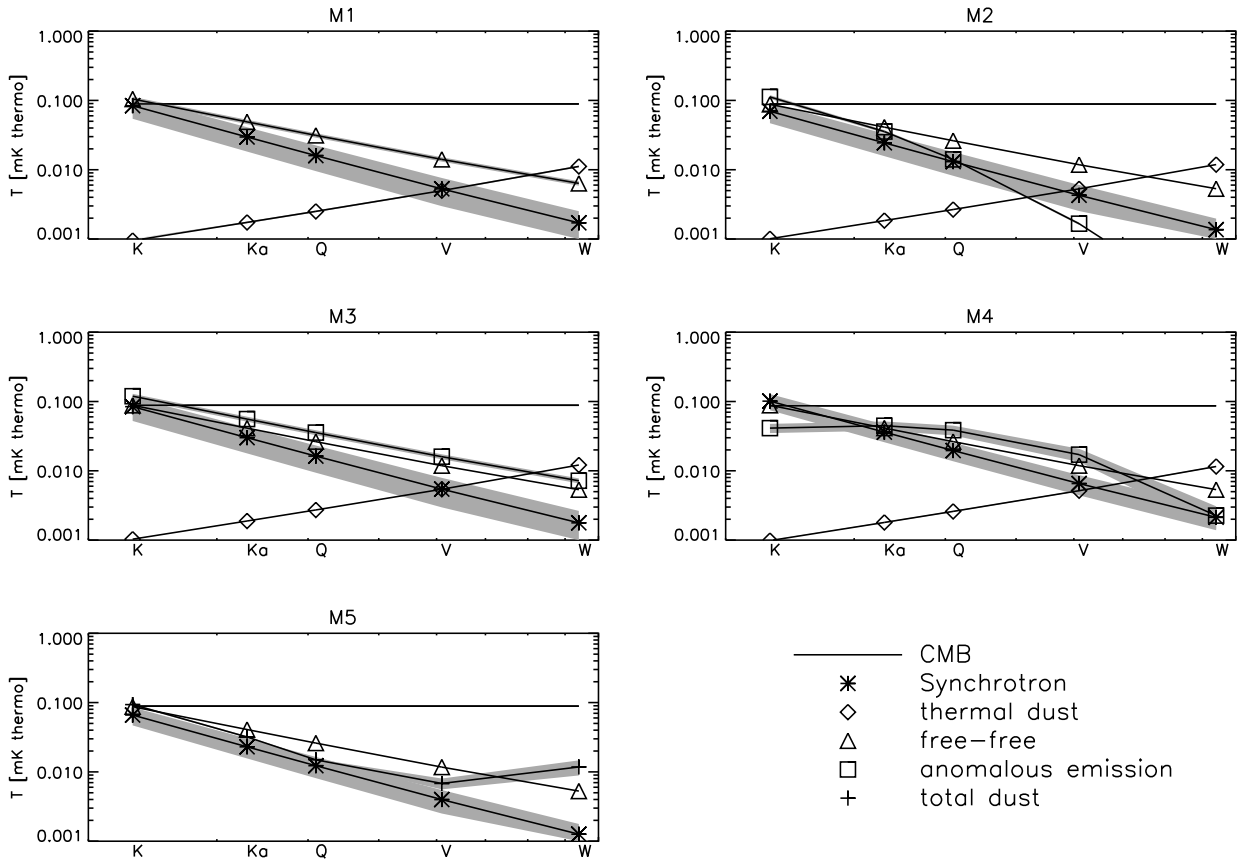}

\caption{Frequency dependence of the rms fluctuations, in
thermodynamic temperature, of the components reconstructed with each
model. The shaded areas, visible only for components whose spectral
parameters are allowed to vary, show the dispersion obtained from
the 30 reconstructions per model, and the lines shows the mean
spectra.} \label{fig:scalings}
\end{center}
\end{figure*}

\section{Reconstructed components} \label{sec:sep}
In Fig.~\ref{fig:scalings} we show the rms in the
WMAP bands of the components reconstructed using each model. Only
pixels outside the apodized mask of Fig.~\ref{fig:apo0} have been
taken into account.
%
%

%
%
On this plot, the CMB component shows tiny variations from one
iteration to another (the shaded regions are too narrow to be
visible) and also from a model to another, suggesting that the
choice of the foreground model, among those considered here, has
little influence on the CMB reconstruction. This will be better
discussed in the next section, where we will focus our analysis on
the CMB component. Variations are very small also for the
reconstructed thermal dust and free-free components. This is due to
the fixed frequency scaling and to the use of prior maps.

As expected, we have a substantial spread in the reconstructed
synchrotron component, reflecting the wide range of synchrotron
spectral indices used for the separation. This component is found to
be generally sub-dominant compared to the free-free, except,
possibly, in the K band, in
agreement with the results by \cite{hinshaw2006}. 



The results on the anomalous emission
components are also quite stable. If we force this component to
have a power-law spectrum (model {\bf M3}) we find that it dominates
over both synchrotron and free-free in all the WMAP channels. 
The anomalous emission for {\bf M4} is found to dominate in the 
Ka, Q and V bands. 
Models {\bf M2} and {\bf M5} are consistent with each other, as the summed
intensity of thermal dust and anomalous emission of model {\bf M2} 
is close to that
of the ``total dust'' emission of model {\bf M5}. Moreover, they 
yield similar intensities for the remaining components. 
As previously stressed, models {\bf M2} and {\bf M5} yield results 
on the anomalous emission which are consistent with \cite{davies2006}.



\subsection{Quality tests}\label{sec:compa}

Before proceeding with the analysis of the
results we try to evaluate the goodness of the
decomposition for each model. A standard method is to analyze the
residual between the data and the reconstructed sources combined by
means of the estimated mixing matrix:
\begin{equation}
 \mathbf{r}\equiv \mathbf{x}- \mathbf{\hat{H}}\mathbf{\hat s}\label{residuals}.
\end{equation}

The analysis of the residuals allows to check if, given a certain 
mixing matrix, our algorithm
 succeeds in finding a plausible decomposition of the data.
In the case of a perfect reconstruction of the channels, $\mathbf{r}$ 
contains maps
of pure noise, characterized by Gaussian statistics.
Foreground residuals bring in non-Gaussianities. 
A visual inspection of the residuals show that 
some non-Gaussian features are present in the maps. 
In particular, we can discern traces of anisotropic noise, point
sources and diffuse emissions, particularly at low Galactic
latitudes. The first two features are related to effects that are
not accounted for in our analysis. The latter is more interesting,
because it also reflects our modeling and estimation errors.

\begin{figure}
\begin{center}
\includegraphics[width=4.cm]{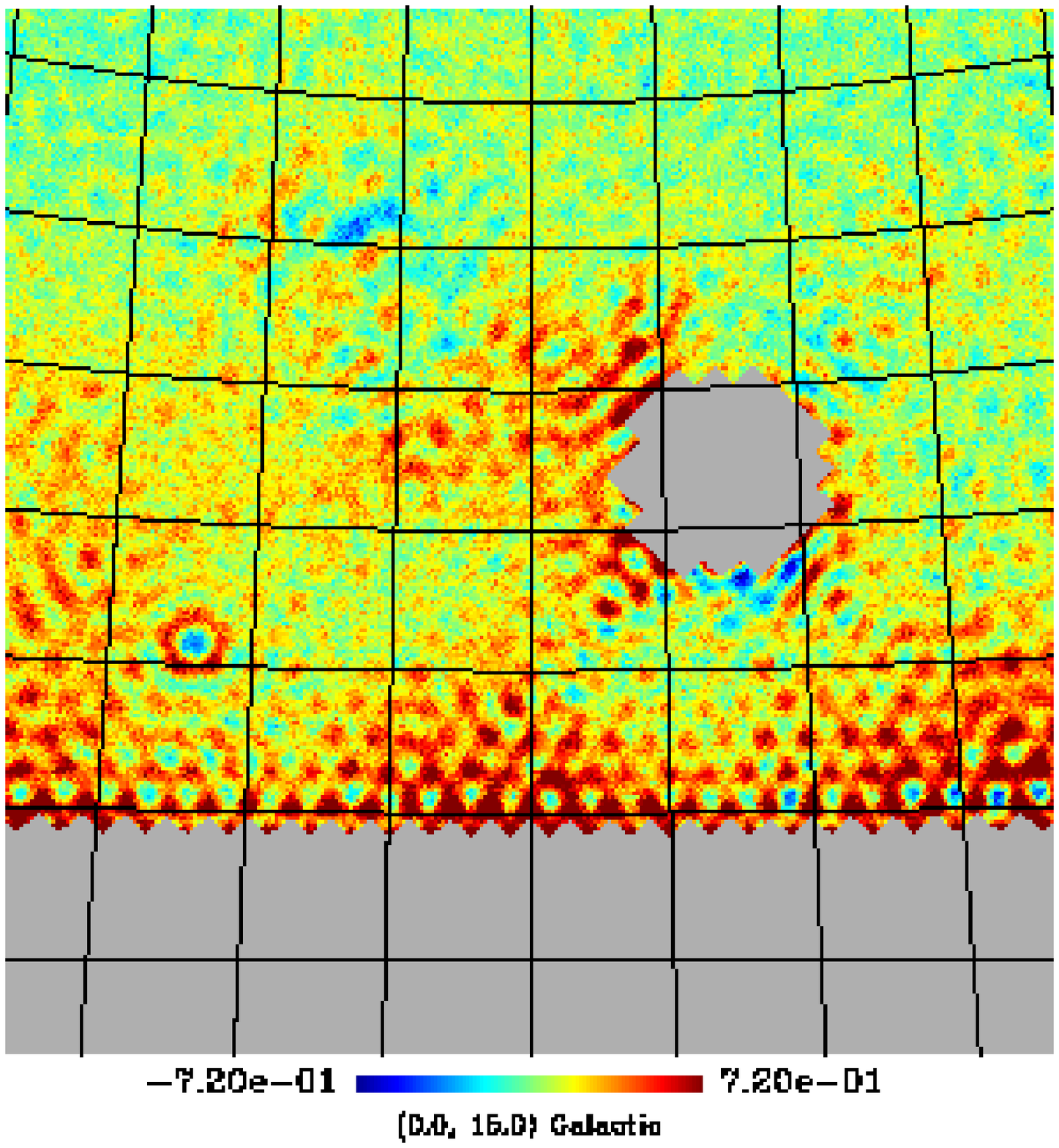}
\includegraphics[width=4.cm]{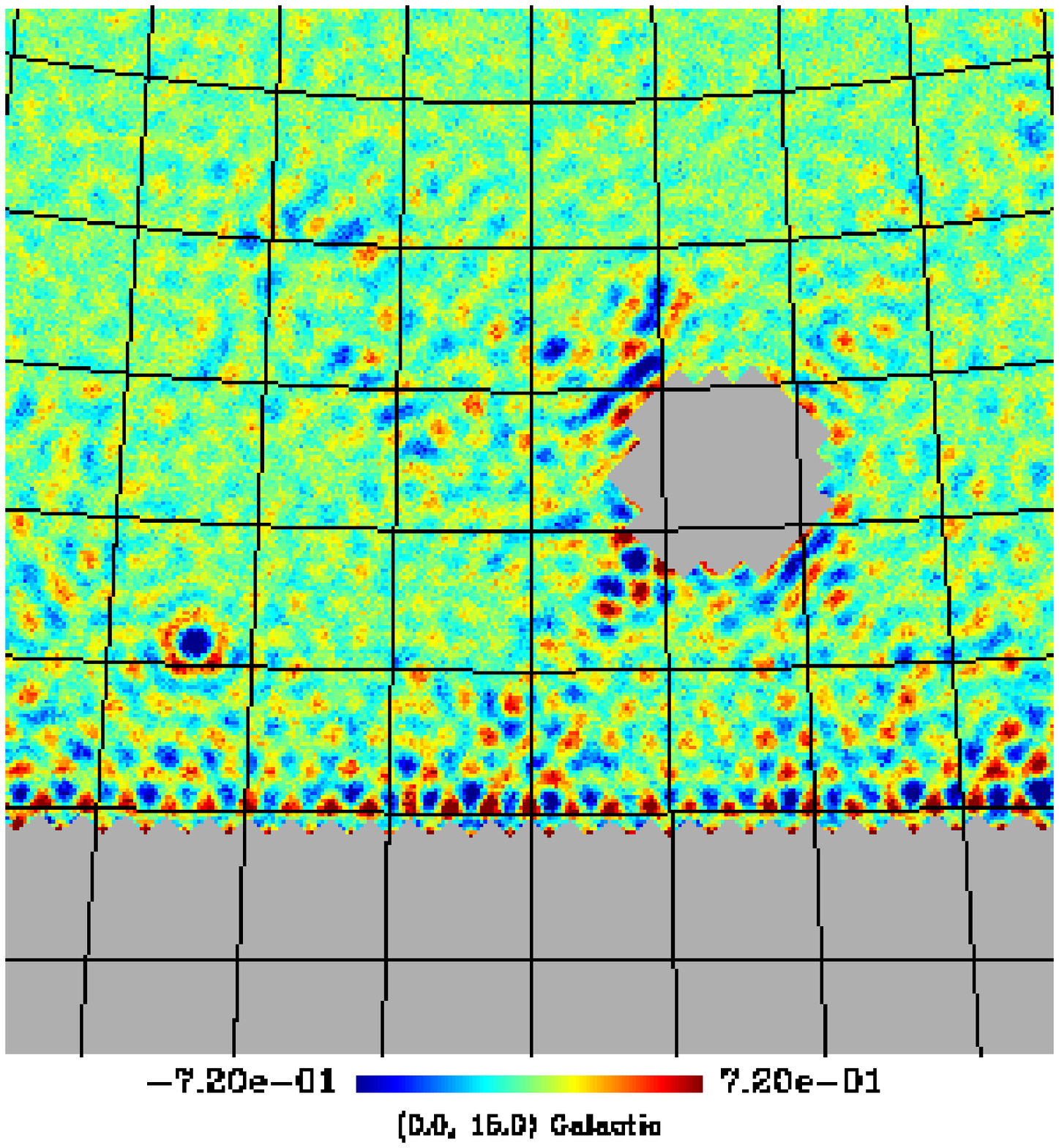}
\includegraphics[width=4.cm]{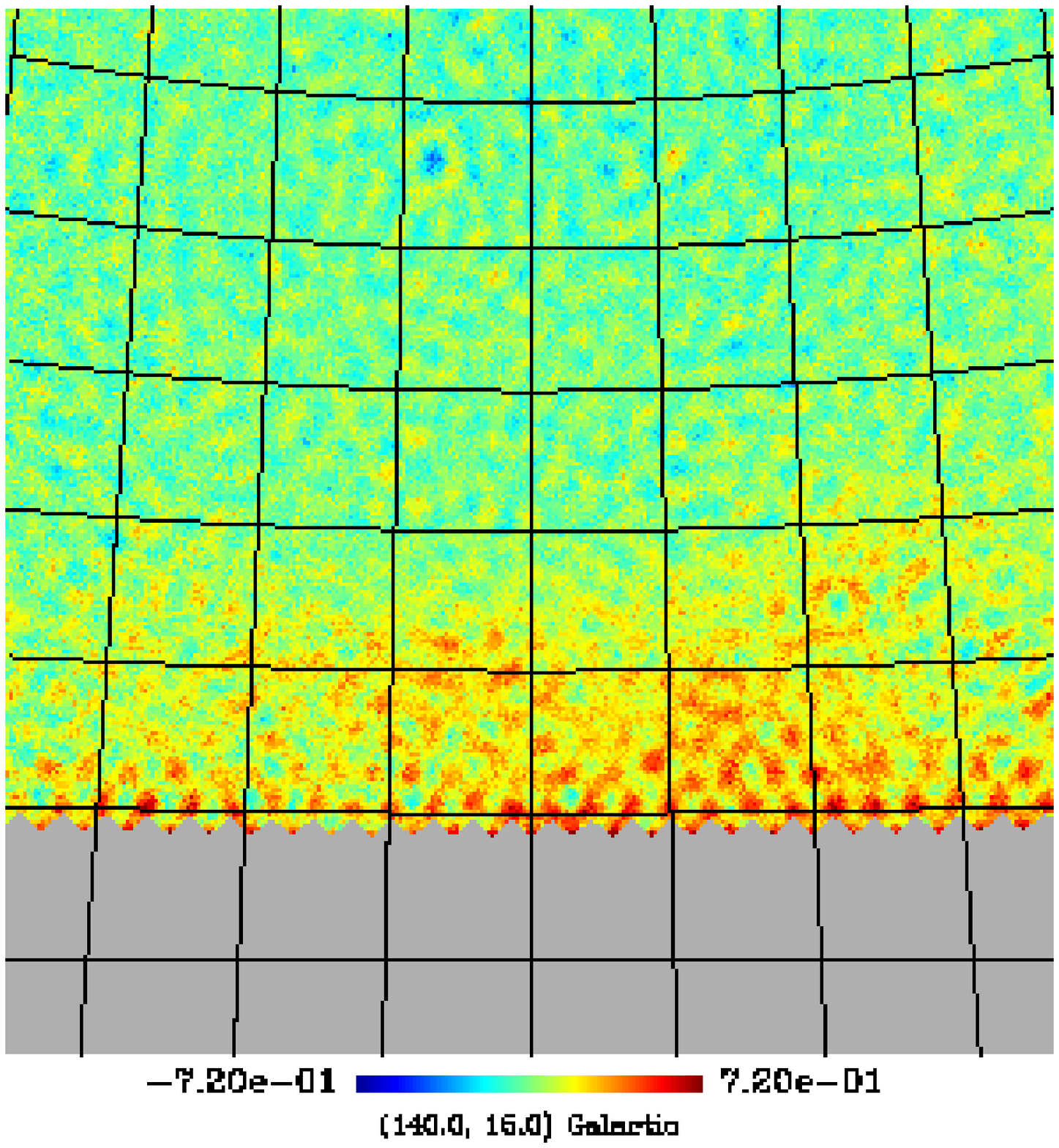}
\includegraphics[width=4.cm]{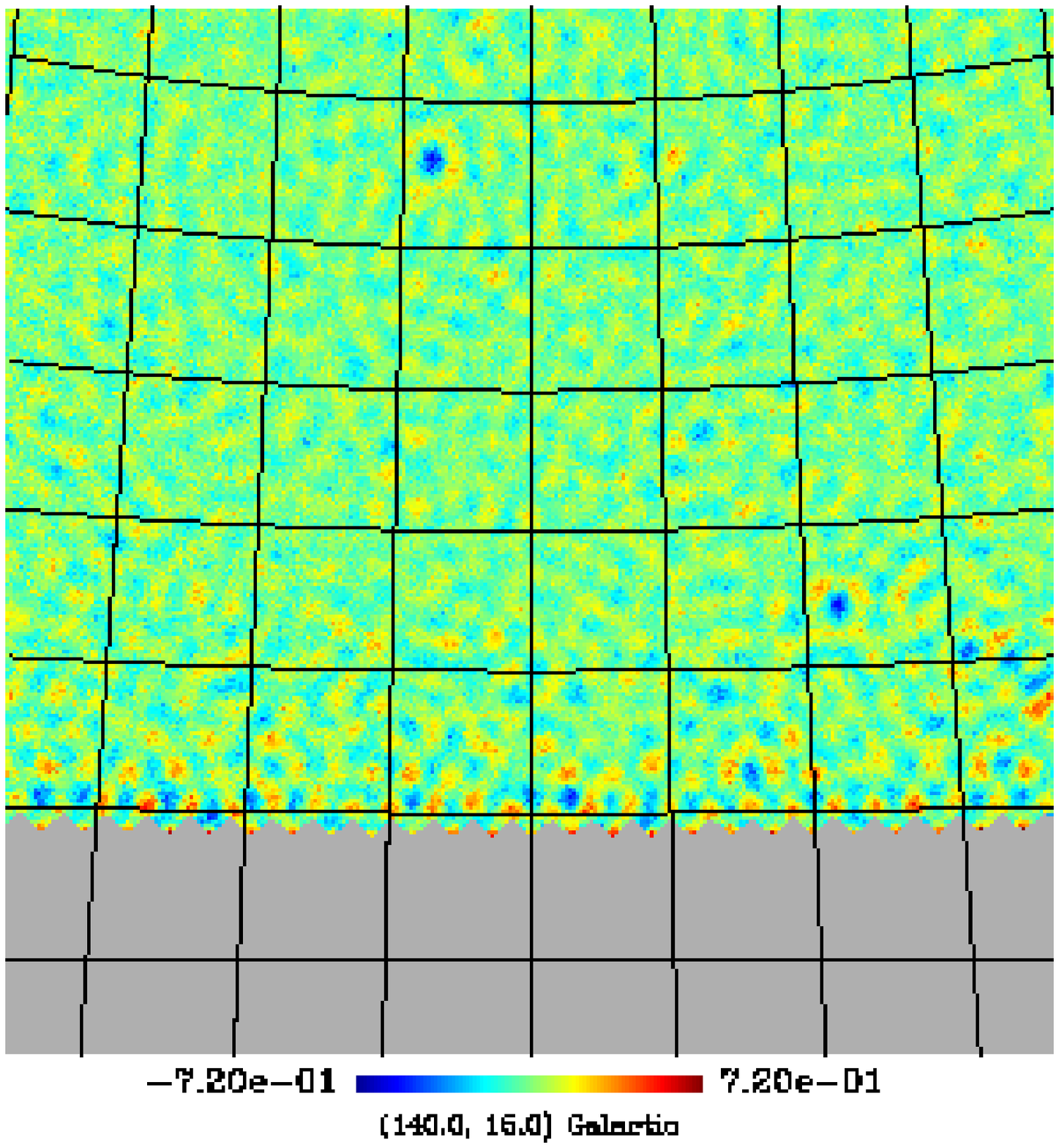}
\caption{Residual at K band for model {\bf M1} (left) and {\bf M2} (righ) assuming $\beta_s \sim 3.$ in two regions of $\sim$ 40X40 degrees. Central coordinates of the patches are $l=0^{\circ}$, $b=15^\circ$ (up) and $l=140^{\circ}$, $b=15^\circ$ (down). Units are mK thermodynamic.}
\label{fig:respatches}
\end{center}
\end{figure}

Each model shows the worst contamination in the K band, even if 
models including the anomalous emission 
are generally more clean (see for example the comparison 
between {\bf M1} and {\bf M2} in Fig. \ref{fig:respatches}).
The ``residual excess'' at low latitudes then decreases at 
higher frequencies. 
The models {\bf M1} and {\bf M4}
are generally worse; however this strongly depends on the
synchrotron spectral index used for each separation.

More quantitative analyses of the residuals (power spectra, 
standardized moments of the distributions, etc.) mainly confirm these 
general findings, but turned out to be not 
very effective in discriminating among the
models, as the results are comparable. 

This means that, in our case, each model allows a
satisfactory decomposition of the data,
 at least for a subset of synchrotron spectral indices. 
Neverthless, some of the models could still give uncorrect descriptions 
of the data.
The analysis of the residuals, infact, is not informative on 
the reliability of the individual reconstructed components. 
For example, they may be affected by
aliasing, which does not show up in the residuals. 
To investigate this possibility, we applied other 
quality tests directly based on the reconstructed components.


\subsubsection{Correlation among the components}\label{sec:corre}


Independent observations have highlighted positive correlations of
varying strength among synchrotron, free-free and dust emissions,
ensuing from the physical processes producing them. Therefore, the
reliability of the recovered components can be tested by
investigating their mutual correlations. We have worked with pixels of area $0.84\,\hbox{deg}^2$ (Nside = 64). We trimmed all pixels pre-processed through the
apodized mask of Fig. \ref{fig:apo0} as well pixels within a radius
of $1^\circ$ around each point source listed in the WMAP three-year
catalogue. We also excluded a region with a radius around $15^\circ$ around the Gum Nebula, as in this region the separations
are always very noisy. High Galactic latitude regions ($|b|>
50^\circ$), where foregrounds are weak and therefore recovered with
low $S/N$ ratios have also been excluded from the analysis. Around $3 \cdot 10^4$ pixels then
remain. 

For each model we have computed the correlation
coefficients, $r$, among the recovered Galactic components for each of
thirty separations.
Whenever one of those coefficients was found to be negative,
the corresponding separation was discarded. The percentages of
discarded separations for each model are reported in Table
\ref{tab:flag}. They are generally low ($\leq$ 10\%), except in the
case of {\bf M3} (30\%).

All the models including the anomalous component feature very strong
positive correlations of it with thermal dust ($r \sim 0.8$),
somewhat less strong but still highly significant positive
correlations with synchrotron ($r \sim 0.6$), and weak positive
correlations with free-free ($r \sim 0.25$).

\begin{table}
\caption{Percentage of discarded separations because of negative
correlation coefficients.}
 \label{tab:flag}
\begin{tabular}{|c|c|}
\hline
model&discarded separations\\
\hline
{\bf M1}&0\%\\
{\bf M2}&6\%\\
{\bf M3}&30\%\\
{\bf M4}&10\%\\
{\bf M5}&3\%\\
\hline
\end{tabular}
\end{table}









\begin{figure}
\begin{center}
\includegraphics[width=5.5cm,angle=90]{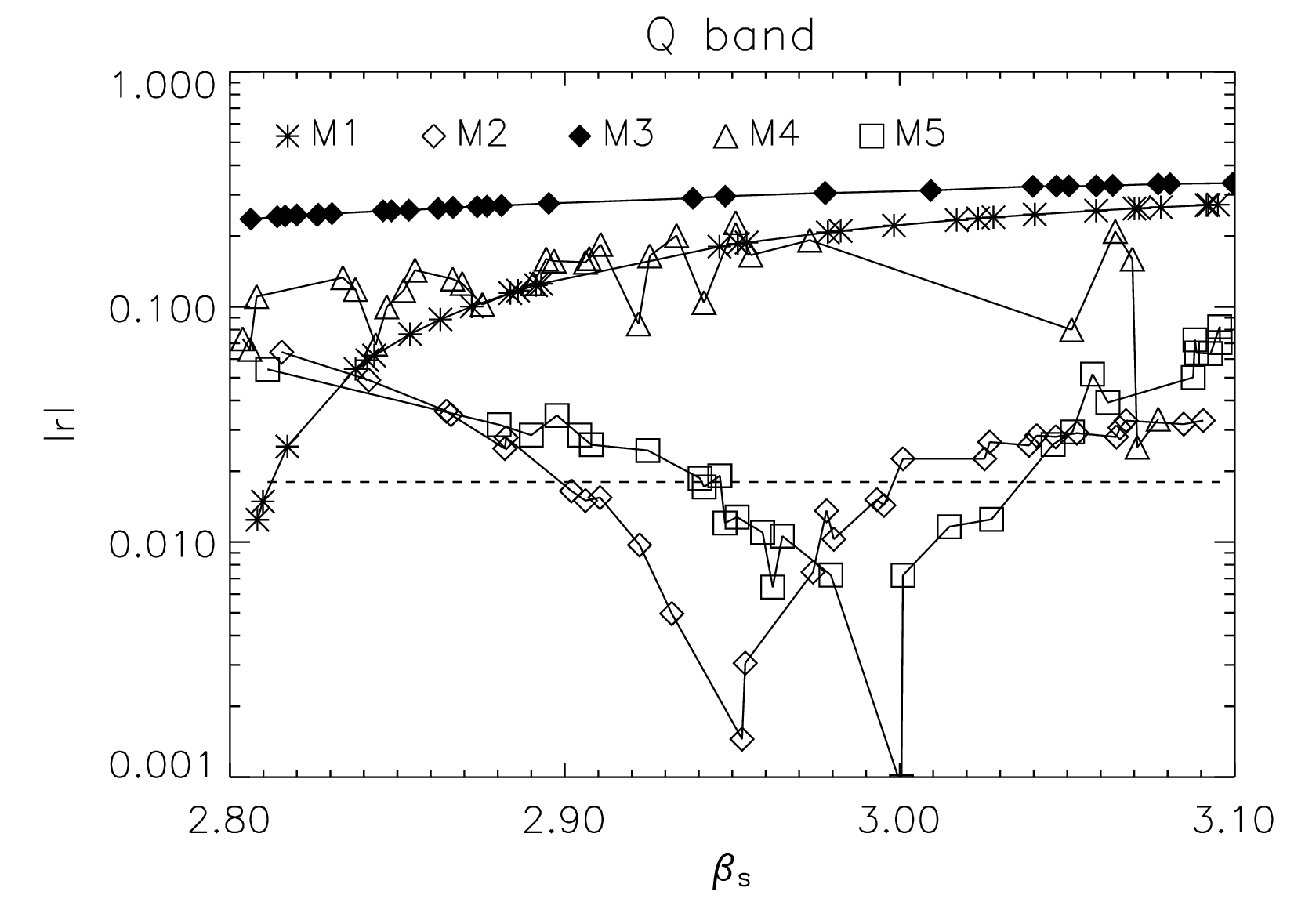}
\includegraphics[width=5.5cm,angle=90]{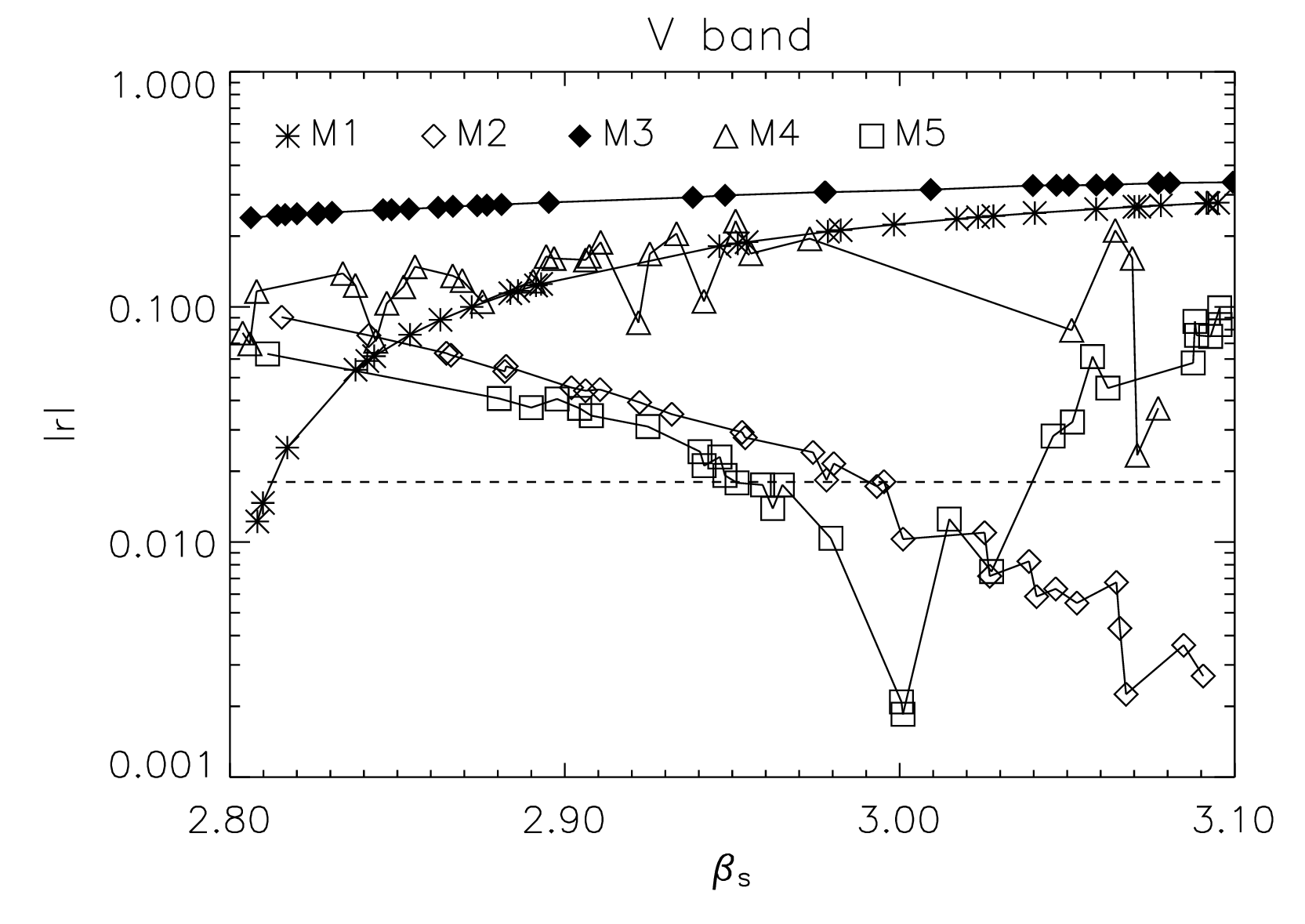}

\caption{Absolute values of the correlation coefficient between the
recovered CMB and the total foreground map for the Q (up) and V
(down) bands as a function of the synchrotron spectral index. Values
above dashed line correspond to correlations formally significant at
 $\geq 3\sigma$.} \label{fig:QVcor}
\end{center}
\end{figure}

For the large pixel number, $N\simeq 3\times 10^4$, we are dealing
with the correlation coefficient, $r$, for samples of uncorrelated
quantities has a normal distribution with zero mean and
$\sigma_r=1/\sqrt{N}$. CMB and Galactic foregrounds are
intrinsically uncorrelated; therefore, the presence of a
statistically significant correlation among them indicates a poor
component separation. Thus, a good criterion to evaluate the quality
of the separations is to compute the correlation coefficient between
the reconstructed CMB and foreground maps. The latter have been
obtained as the sum of the reconstructed components scaled to the Q
and V bands. As in these bands the foreground contamination has a
minimum, those correlations are expected to be low and, hopefully,
not statistically significant. Once computed for each separation and
each model, the correlation coefficients can be used as figures of
merit to compare different models and, for a given model, to
identify the best separations.

In Figure \ref{fig:QVcor} we show the absolute values of the
correlation coefficients as a function of the assumed synchrotron
spectral index. The horizontal dashed line is the 3$\sigma$
significance level.

Model {\bf M3} always features highly significant correlations
($r>0.1$) between the CMB and the reconstructed foreground
contribution in both bands, much stronger than found for the other
models. This result, together with the high number of samples we had
to discard due to aliasing effects, indicates that this model is not
able to correctly describe the data. It was therefore discarded
together with model {\bf M4} which also yields values of $|r|$
always well above the 3$\sigma$ limit, although somewhat below those
found for {\bf M3}.

In the case of models {\bf M1}, {\bf M2} and {\bf M5}, the test
shows that the correlations are not statistically  significant for
well defined ranges of the synchrotron spectral index. Model {\bf
M1} requires relatively low values of $\beta_s$, below that
expected, in the WMAP frequency range, from the locally measured
energy spectrum of relativistic electrons which would yield
$\beta_s\simeq 3$ \citep{banday1990, banday1991}, as indeed found
for models {\bf M2} and {\bf M5}.


We have built reference maps of each reconstructed component for
each of the 3 surviving models by averaging those reconstructions
with foreground--CMB correlations at less than $3\sigma$ level. For model
{\bf M2} we find somewhat different allowed ranges of $\beta_s$ for
the Q and the V bands, and we have kept the reconstructions in the
union of the 2 ranges, i.e. with $2.9 \lsim \beta_s\lsim 3.1$.  In
Table ~\ref{tab:corr_cmb} we list the correlation coefficients of
the reference CMB map for each model with the foreground templates.
The correlations computed with the minimal mask are formally statistically significant, indicating
the presence of residual foreground contamination. The major
contributions to the correlation coefficients come from regions
close to the Galactic plane. If we adopt the WMAP kp2 mask, in place
of our minimal mask, the correlation coefficients decrease, sometimes going close 
to or below the $3\sigma$ significance level, especially for free-free. In Table~\ref{tab:corr_tmp} we
show the correlations of the reconstructed synchrotron, dust and
free-free components with the corresponding templates. The match
between the prior and reconstructed maps is good but not perfect, as
expected given the spatial variations of the spectral properties.
Finally, Table~\ref{tab:corr_sd+tmp} gives the correlation
coefficients between the anomalous emission recovered with model
{\bf M2}, and the thermal dust plus anomalous emission recovered
with model {\bf M5} (which treats them as a single component with a
complex spectrum), and the foreground templates. Interestingly, in
the case of model {\bf M2}, which does not impose a priori any
correlation between thermal dust and anomalous emission, we find a
tight, but not perfect, correlation among the 2 components. The
anomalous emission also correlates strongly with synchrotron, and
more weakly with free-free.

\begin{figure}
\begin{center}
\includegraphics[width=5.5cm,angle=90]{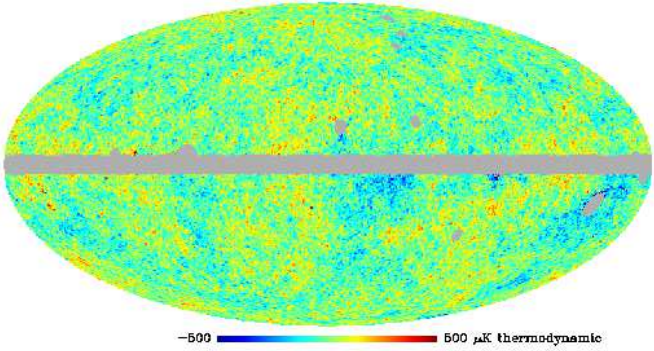}

\caption{CMB map reconstructed with model {\bf M2}.}

\label{fig:cmbmap}
\end{center}
\end{figure}

\begin{table}
\caption{Correlation coefficients of the reconstructed CMB
components  with the foreground templates: minimal mask, (kp2 mask)} \label{tab:corr_cmb}
\begin{tabular}{|c|c|c|c|}
\hline
template&{\bf M1}&{\bf M2}&{\bf M5}\\
\hline
thermal dust&0.027 (0.015)&0.029 (0.023)&0.025 (-0.023)\\
synchrotron&-0.054 (-0.046)&-0.016 (-0.012)&-0.015 (-0.010)\\
free-free&-0.035 (-0.021)&-0.035 (-0.0003)&-0.044  (-0.011)\\
\hline

\end{tabular}
\end{table}


\begin{table}
\caption{Correlation coefficients of the synchrotron, thermal dust
and free-free maps with the corresponding templates. We recall that
for {\bf M5} the dust component includes both the thermal and the
anomalous contributions.} \label{tab:corr_tmp}
\begin{tabular}{|c|c|c|c|}
\hline
template&{\bf M1}&{\bf M2}&{\bf M5}\\
\hline
thermal dust&0.966&0.967&0.883\\
synchrotron&0.865&0.858&0.851\\
free-free&0.957&0.984&0.984\\
\hline

\end{tabular}

\end{table}
\begin{table}
\caption{Correlation coefficients of the anomalous emission ({\bf
M2}) and of the total dust component ({\bf M5}) with all the
foreground templates.} \label{tab:corr_sd+tmp}

\begin{tabular}{|c|c|c|}
\hline
template&{\bf M2} &{\bf M5}\\
\hline
thermal dust&0.823&0.883\\
synchrotron&0.664& 0.711\\
free-free&0.294&0.221\\
\hline

\end{tabular}

\end{table}

\begin{figure*}
\begin{center}
\includegraphics[width=6.cm,angle=90]{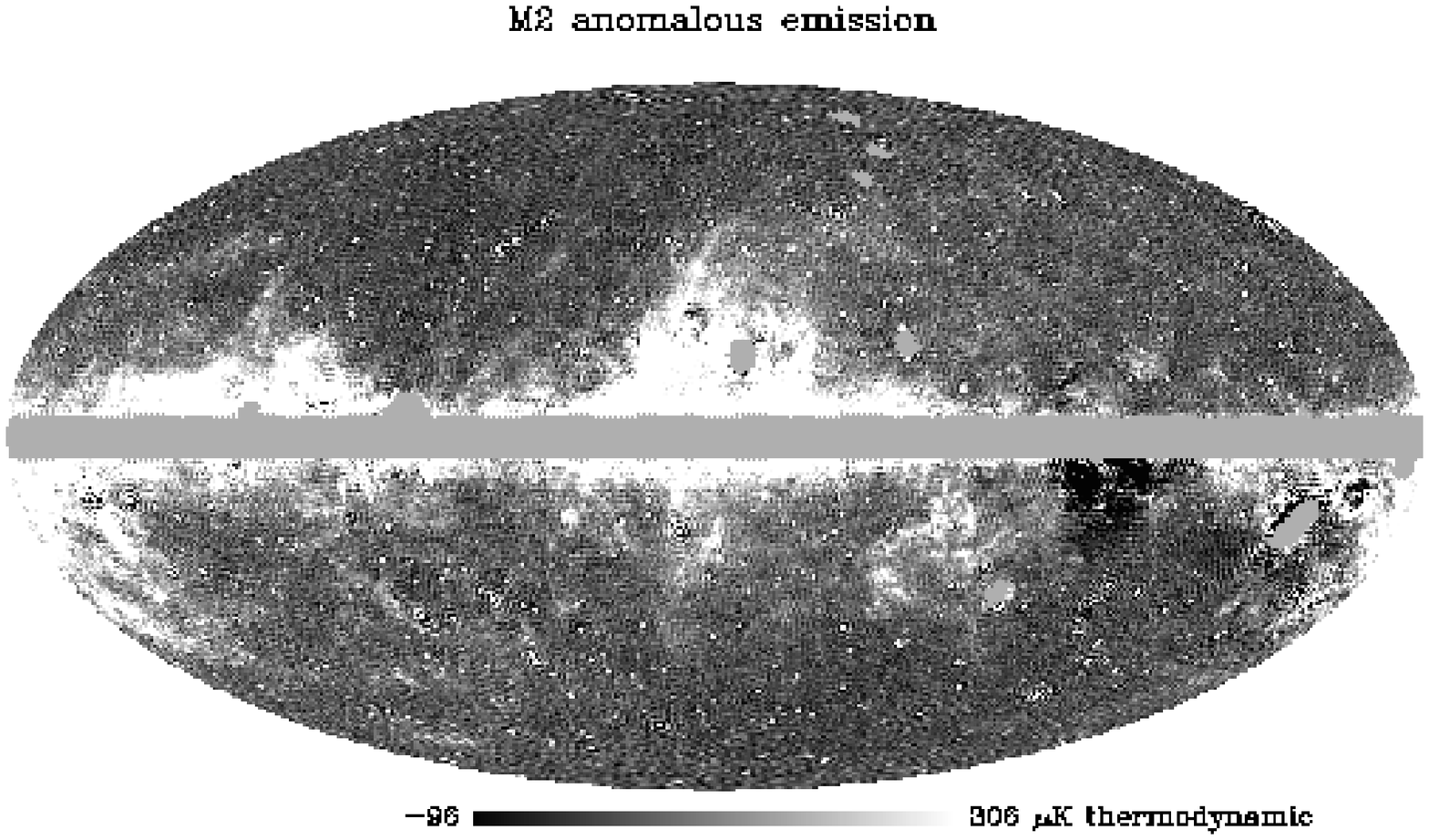}
\includegraphics[width=6.cm,angle=90]{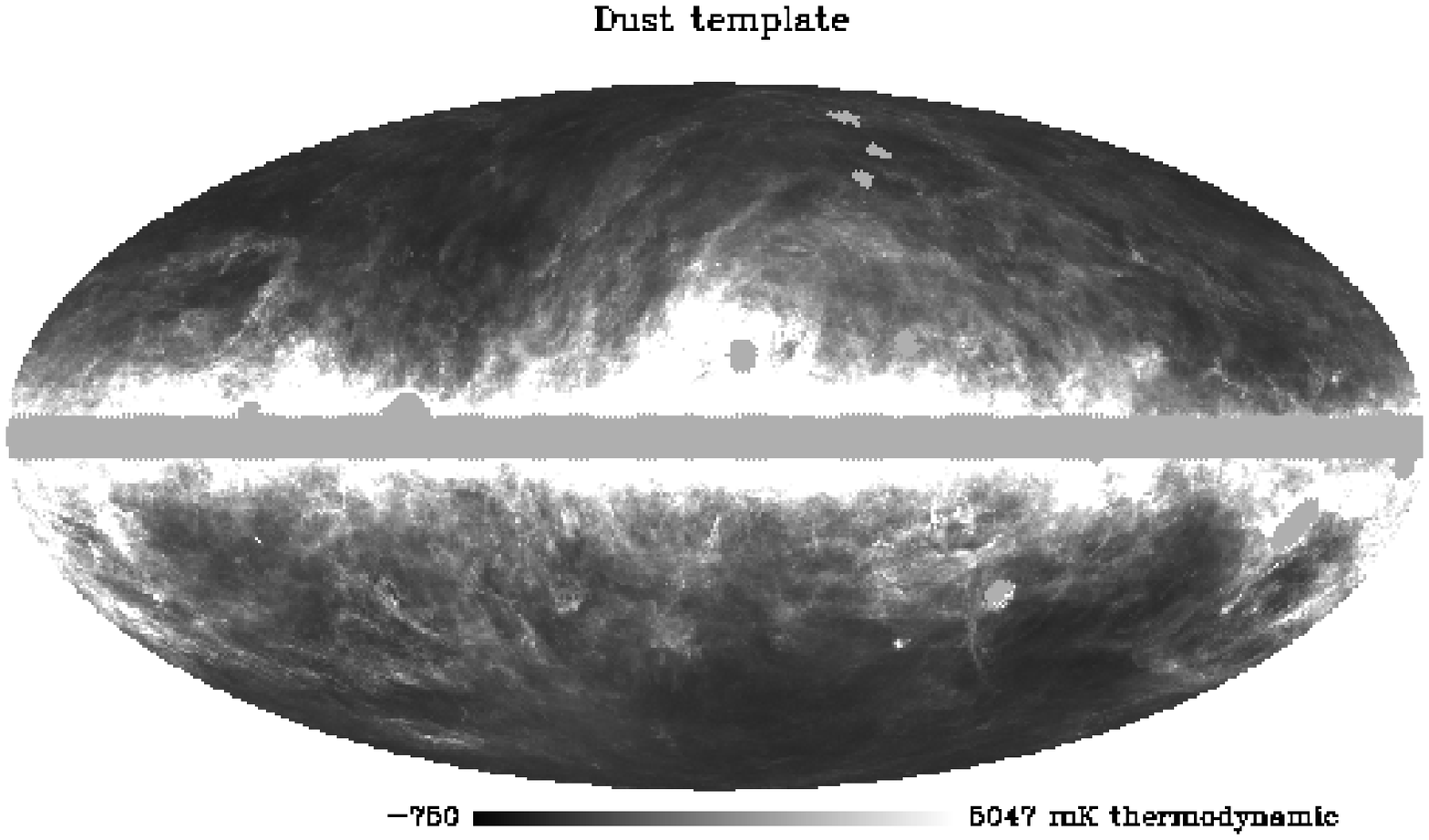}
\includegraphics[width=6.cm,angle=90]{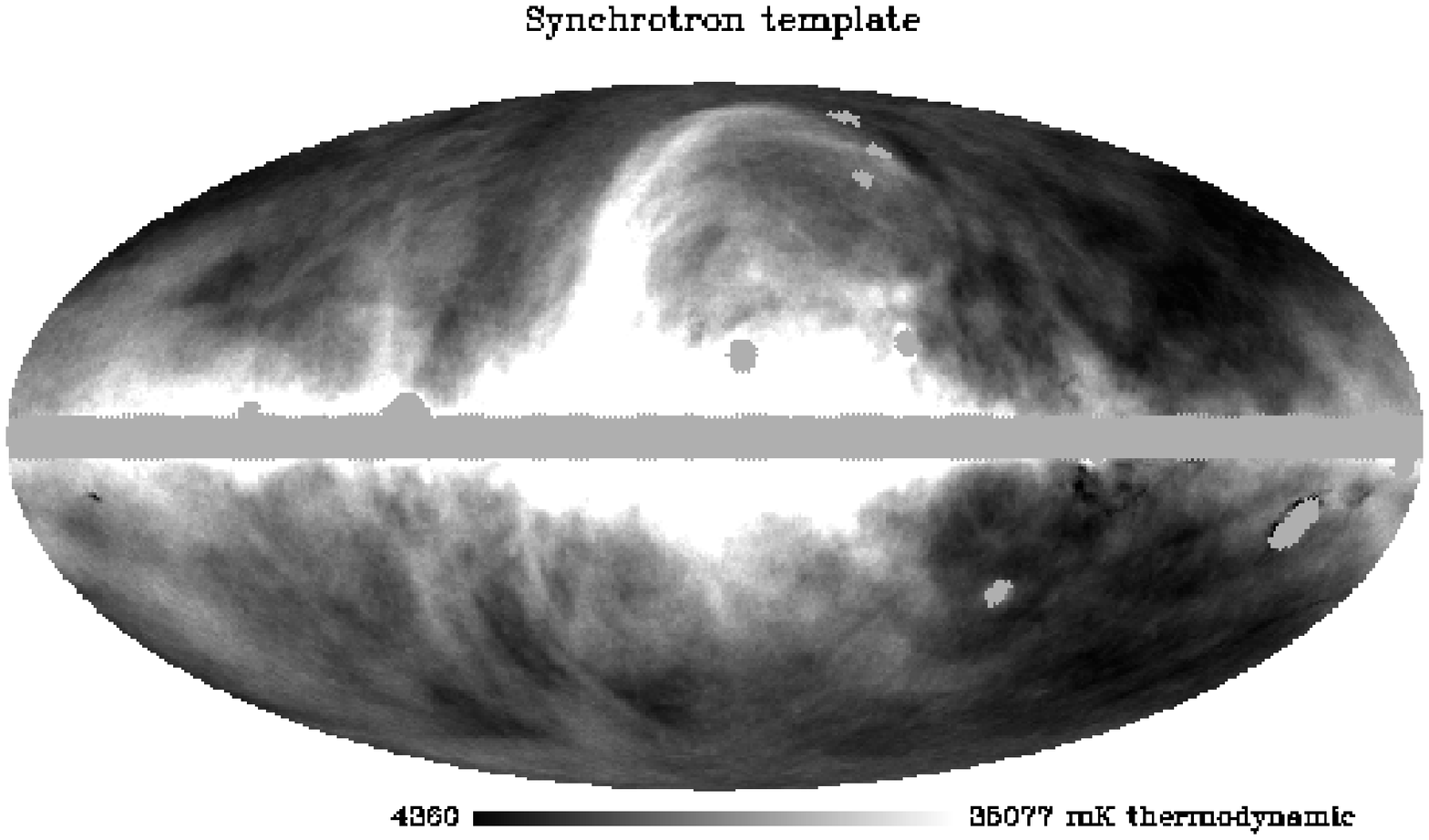}

\caption{Map of the anomalous emission recovered with model {\bf M2}
in the K band compared with the dust (850 GHz) and the synchrotron
(408 MHz) templates. Minimum and maximum value of the colour scales are given by the mean of the map $\pm$ its standard deviation.}

\label{fig:anomalous}
\end{center}
\end{figure*}
\begin{figure*}
\begin{center}
\includegraphics[width=5cm]{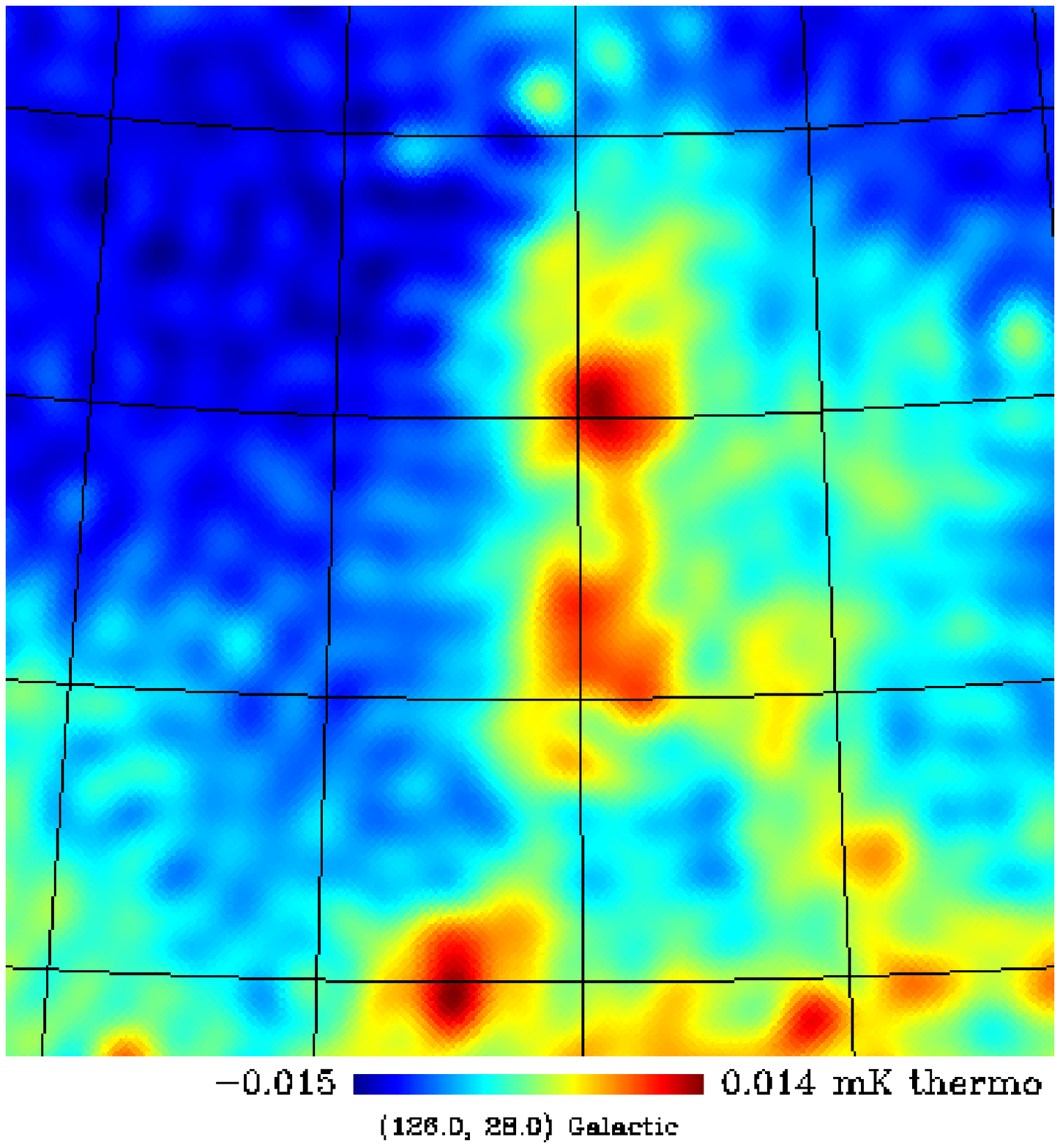}
\includegraphics[width=5cm]{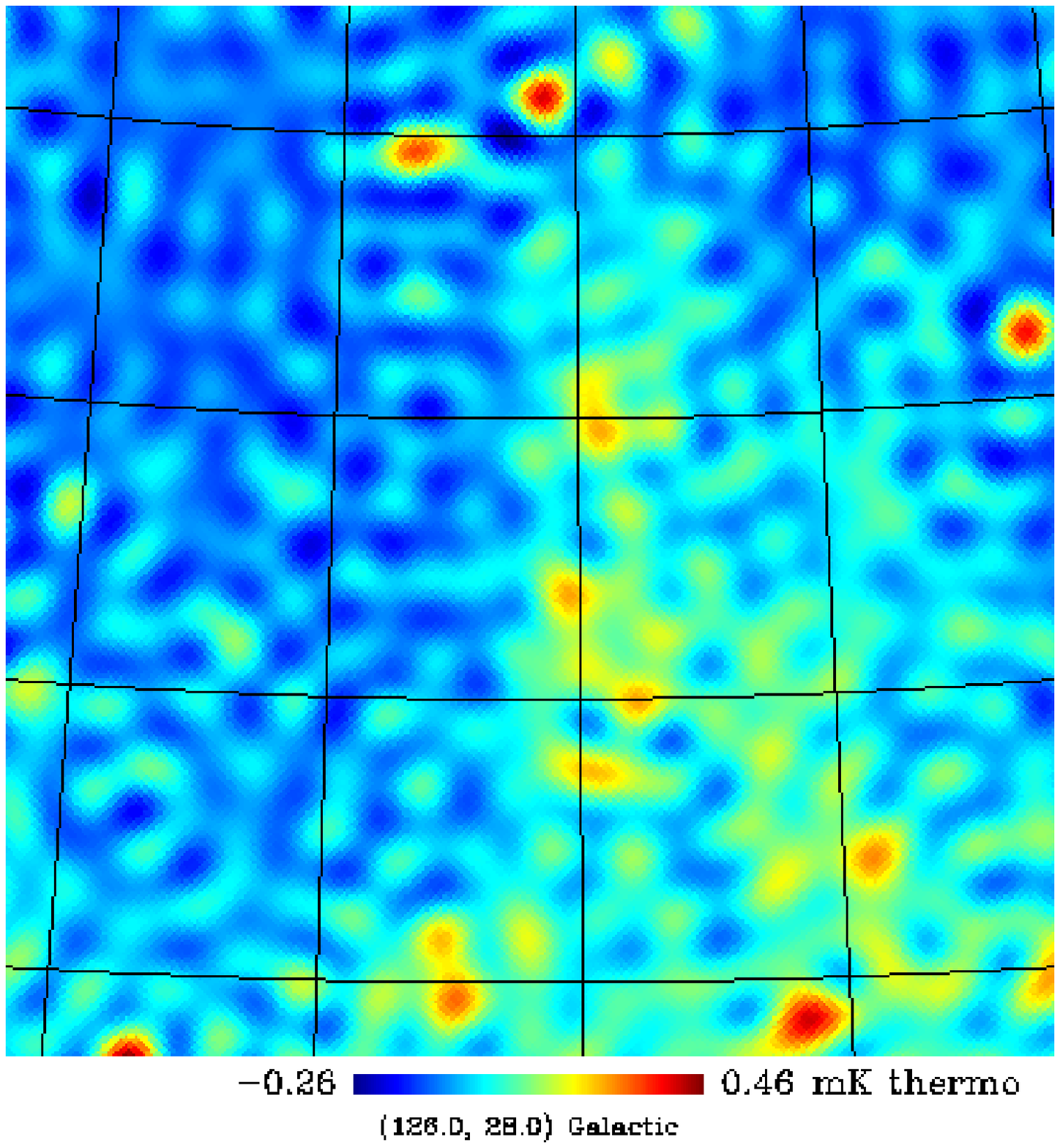}
\caption{Recovered dust (left) and anomalous emission (righ) in Region 6 of Davies et al.(2006). Fluxes (mK thermodynamic) are at W and K band respectively.}
\label{fig:zonedavies}
\end{center}
\end{figure*}

The final map of the anomalous component recovered with model {\bf
M2} is shown in Fig.~\ref{fig:anomalous}, where the dust and
synchrotron templates are also shown for comparison. Traces of
imperfect component separation can be discerned by eye:  negative
imprints in correspondence of the Gum Nebula, of the Orion region
and of the North Polar Spur, as well as several point-like sources.
Nevertheless, this may be the first, albeit preliminary all-sky map
of this component. The similarity with the thermal dust map is
evident; we recall that no prior correlation of the anomalous
emission with thermal dust was imposed. Differences with the spatial
distribution of synchrotron are also clear: the anomalous emission
is relatively stronger towards the outskirts of the Galaxy, where
the first application of CCA found flat ``synchrotron'' spectral
indices, and less prominent near the Galactic centre.
In Fig. \ref{fig:zonedavies} we show a detail of the comparison between 
the thermal dust and anomalous emission maps recovered with {\bf M2}. 
The area we show is the Region 6 of \cite{davies2006}, chosen to be
dust dominated and poorly contaminated by free-free and synchrotron.
In agreement with their results, the anomalous emission is detected in this region, and appears to be dust-correlated.



\section{The CMB power spectrum} \label{sec:cmb}

\begin{figure}
\begin{center}
\includegraphics[width=8cm]{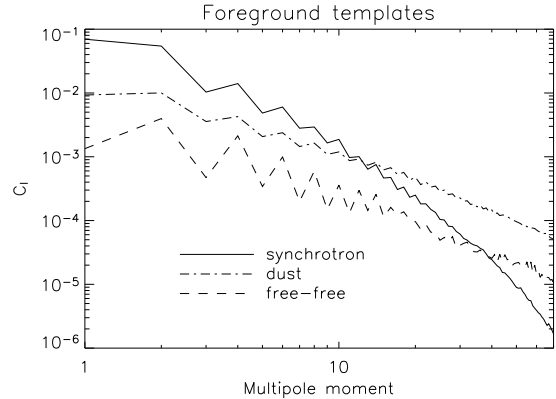}

\caption{Full-sky power spectra of the foreground templates at low
multipoles (arbitrary normalizations)} \label{fig:templates}
\end{center}
\end{figure}


The next question we wish to address is: to what extent do the
different foreground models affect the estimates of the CMB power
spectrum on large scales, where possible inconsistencies with
predictions of the standard cosmological paradigm have been
reported?

We will focus our analysis on $l\le 70$, where the effect of
fluctuations due to unresolved point sources, non included in our
study, can be neglected. On large scales, however, the CMB maps
reconstructed through matrix inversion of the WMAP data complemented
with the three template foreground maps, keep trace of the spurious
structure due to latter maps being mosaics of observations. This is
clearly visible in Fig.~\ref{fig:templates} showing the power
spectra of the templates at low multipoles. While for $l\gsim 10$
the power spectra are smooth, at lower multipoles there are
conspicuous spikes, particularly for the free-free and the
synchrotron. To avoid this problem, the CMB power spectrum for $l <
10$ was computed using only the WMAP data for the Wiener Filter
reconstruction of the CMB, still using the same ranges of optimal
spectral parameters for each model. For larger $l$'s the inversion
is carried out including the foreground templates. The error
analysis, carried out as described in \cite{bonaldi2006}, confirms
that this approach significantly decreases the errors on the power
spectrum at low $l$'s. The power spectra obtained with the two
methods are very close to each other for $l \simeq 10$, so that they
smoothly join and the choice of the boundary multipole needs not to
be fine tuned.





In these calculations we used two sky masks. The first one (minimal
mask) excludes all pixels having non unitary value in the apodized
mask used to preprocess the data (Figure~\ref{fig:apo0}); the cut
region amounts to $\sim 10\%$ of the sky. The second is the previous
one multiplied by the WMAP kp2 mask; it excludes $\sim 20\%$ of the
sky. The latter is less liable to residual foreground contamination,
while the former is less liable to biases due to incomplete sky
coverage.  The power spectrum was binned with the same scheme
adopted by the WMAP team and we applied the MASTER approach
\citep{hivon2002} to get an unbiased estimate.

\begin{figure}
\begin{center}
\includegraphics[width=5.5cm,angle=90]{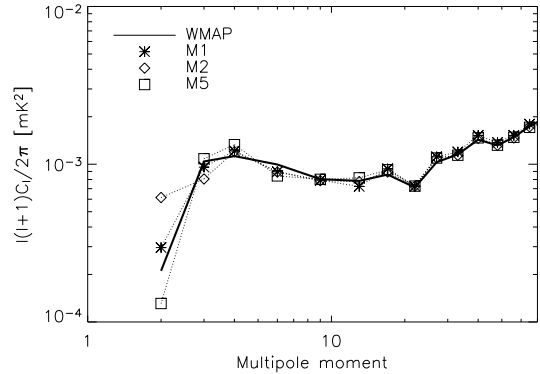}
\caption{Binned CMB power spectra obtained from all models compared
to the WMAP three-year power spectrum} \label{fig:cmbps}
\end{center}
\end{figure}

The results for each model are shown in Fig.~\ref{fig:cmbps}, where
we have plotted the power spectra computed with the minimal mask for
$l \leq 8$, those computed with the other mask for higher
multipoles. Again the results are weakly dependent on the choice of
the mask, but the minimal mask yields smaller oscillations at low
$l$'s. We find generally a good agreement with the WMAP power
spectrum, shown by the solid line. There are however significant
differences among the various models at the lowest multipoles, especially evident for $l=2$.
Regardless to which are the actual models we are exploiting, this result 
shows that the hypotheses on the Galactic components made to perform 
component separation can be painful for the CMB power spectrum at larger 
scales. Thus, big error bars are 
required to take into account for possible biases at these multipoles. 
As an estimate of uncertainties associated to
foreground modeling, we can take the spread of our three CMB power spectra.

\begin{figure}
\begin{center}
\includegraphics[width=5.5cm,angle=90]{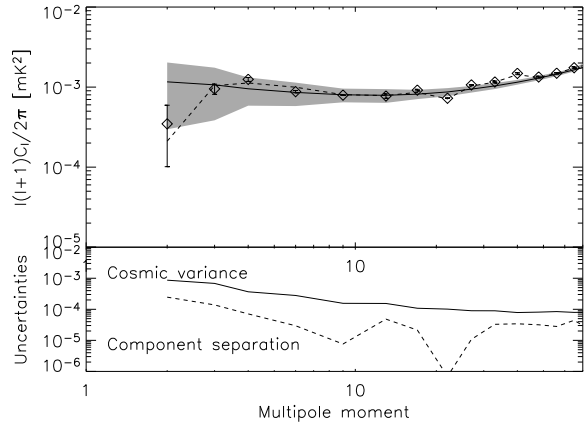}
\caption{Upper panel: best fit $\Lambda CDM$ model(smooth solid
line) compared to the WMAP CMB power spectrum (dashed line) and to
the mean final power spectrum resulting from our analysis (open
diamonds; the error bars show the spread of results for different
models). The shaded area shows the cosmic variance. Lower panel:
uncertainties associated to foreground modeling (dotted line)
compared to the cosmic variance (solid line).} \label{fig:lcdm}
\end{center}
\end{figure}

In the upper panel of Fig.~\ref{fig:lcdm} we compare our results
with the WMAP three year power spectrum and with the best fit
$\Lambda$CDM model based on WMAP data only \citep{spergel2006}. The
mean quadrupole moment coming out from our analysis is higher than
the WMAP estimate, and the difference with the prediction of the
$\Lambda$CDM model is within the uncertainty due to cosmic variance.
The spread of estimates of the quadrupole moment from different models is 
of about $\pm 200 \mu\hbox{K}^2$.

\begin{figure}
\begin{center}
\includegraphics[width=5.5cm,angle=90]{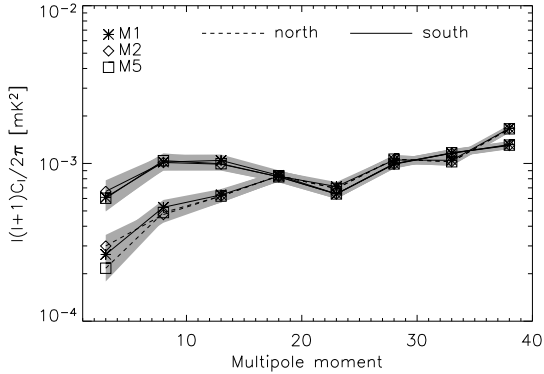}
\caption{North-south asymmetry. The bins adopted have central multipoles
 $l=\{3,8,13,18,23,28,33,38\}$  and equal width $\Delta l=5$}.
\label{fig:ns}
\end{center}
\end{figure}

Summing in quadrature the cosmic variance and the modeling errors,
we find no large scale power spectrum ``anomalies'' significant at
$\ge 1.5 \sigma$, except for the excess power at $l \simeq 40$,
which is significant at $\simeq 4\sigma$. On the other hand, as show
by Fig.~\ref{fig:ns}, the North-South asymmetry at $l \lsim 20$
stands, independently of the adopted foreground model. Those power
spectra, computed for each model for the North and for the South
hemisphere separately, were obtained as previously described, except
for the different binning scheme adopted.

\section{Conclusions}

We have performed the component separation on WMAP maps exploiting
CCA technique to estimate the mixing matrix directly from the
WMAP data, taking also into account the complementary information
from the \cite{haslam1982} 408 GHz map, from the extinction
corrected H$\alpha$ map \citep{dickinson2003}, and from the combined
COBE/IRAS thermal dust maps \citep{finkbeiner1999}. The H$\alpha$
map is exploited also to subtract the free-free contribution, which
is substantial in some regions of the Galactic plane, from the
\cite{haslam1982} map, thus obtaining a pure synchrotron template.

The CCA is particularly useful to deal with possible additional
foreground components, for which little or no prior information
exists. An application assuming that only the canonical Galactic
emissions (synchrotron, free-free, and thermal dust) are present
highlights the widespread presence of a spectrally flat
``synchrotron'' component (Fig.~\ref{fig:isto_M1}), largely
uncorrelated with the synchrotron template (compare
Fig.~\ref{fig:validzones} with the synchrotron template in
Fig.~\ref{fig:anomalous}), suggesting that an additional component
is required.

We have then tested various spectral shapes for such component,
namely a power law (model {\bf M3}) as expected if it is a flat
synchrotron emission due to enhanced magnetic fields in dusty
star-forming regions, or a parabola in the $\log S$--$\log \nu$
plane (model {\bf M2}), or a grey body, as proposed by Tegmark et
al. (2000; model {\bf M4}). In all these cases, the spatial
distribution of this component was left totally unconstrained, but
we have also tested the possibility that it is distributed exactly
as thermal dust (model {\bf M5}).


The CMB maps yielded by models {\bf M3} and  {\bf M4} turned out to
be substantially correlated with foreground maps, indicating a
unacceptably large residual contamination. These models were
therefore discarded. Conversely, in the case of models {\bf M2} and
{\bf M5} the correlations were found to be not statistically
significant for values of the synchrotron spectral index, $\beta_s
\simeq 3$, close to the expectations from the slope of the locally
measured energy spectrum of relativistic electrons (Banday \&
Wolfendale 1990, 1991). The data do not allow us to clearly
discriminate among these models; however, some interesting
indications emerge:

\begin{itemize}

\item The additional component turns out to be always tightly
correlated with thermal dust although, if its spatial distribution
is unconstrained, the correlation is not perfect (correlation
coefficient $r \simeq 0.8$). This reminds us the suggestion by
\cite{davies2006} that, if this component is due to spinning dust,
it should be better correlated with the small grains dominating the
mid-IR emission than with the big grains dominating at far-IR to
sub-mm wavelengths. The map of this component yielded by model {\bf
M2} could then constitute its first, albeit preliminary all sky map.

\item The additional component is well correlated with
synchrotron (correlation coefficient $r \simeq 0.6$) and more weakly
correlated with the free-free, although the free-free is strongly
correlated with thermal dust (correlation coefficient $r =0.4-
0.6$). The synchrotron is also well correlated with thermal dust
(correlation coefficient $r = 0.7-0.8$), and more weakly with the
free-free (correlation coefficient $r \simeq 0.2$).


\item If only the standard Galactic emissions (synchrotron, free-free
and thermal dust) are taken into account, we obtain acceptable
separations adopting a flat synchrotron spectral index ($\beta_s
\sim 2.8$). However, this value of $\beta_s$ is not supported by the
CCA analysis and is flatter than expected in the WMAP frequency
range (Banday \& Wolfendale 1990, 1991).

\item The CMB maps we obtain have low foreground contamination even
at low Galactic latitude, as demonstrated by the correlation with
the foreground templates. The inclusion of the anomalous emission
decreases the synchrotron contamination. 

\end{itemize}

We have used the MASTER approach (Hivon et al. 2002) to obtain
unbiased estimates of the CMB power spectrum up to $l=70$ from maps
reconstructed on the basis of each foreground model. The results are
generally consistent with the WMAP ones, although we find a
substantial spread of values for the quadrupole moment. For model
{\bf M2} the quadrupole moment is fully consistent with the
expectation from the ``concordance'' cosmological model. The spread
of power spectrum estimates gives us a measure of uncertainties
associated to foreground modeling. Such uncertainties are smaller
than the cosmic variance, but nevertheless significant, for $l\lsim
30$, and approach it for larger $l$'s. Combining the cosmic variance
with the modeling errors, we find that the quadrupole amplitude is
less than $1\sigma$ below that expected from the standard
$\Lambda$CDM model. Also the other reported deviations from model
predictions are found not to be statistically significant, except
for the excess power at $l\simeq 40$. On the other hand, the
North-South asymmetry is found to be unrelated to foreground models.
The derived CMB power spectra are remarkably stable: even those
yielded by the discarded models are within the range of
Fig.~\ref{fig:lcdm}.

 \bibliographystyle{mn2e}
\bibliography{master}

\end{document}